\documentclass[a4paper,11pt]{article}
\pdfoutput=1 

\usepackage{jheppub} 

\usepackage[T1]{fontenc} 

\usepackage{graphicx}

\usepackage{amsmath}
\usepackage{amsfonts}

\usepackage{hyperref}

\AtBeginDocument{
  \label{CorrectFirstPageLabel}
  
}

\newcommand{\id}{\mathrm{d}}
\newcommand{\eps}{\epsilon}
\newcommand{\define}{\equiv}
\newcommand{\sminus}{\scalebox{0.50}[0.75]{\( \, - \)}}

\title{On Epsilon Factorized Differential Equations for Elliptic Feynman Integrals}

\author[a]{Hjalte Frellesvig}

\affiliation[a]{Niels Bohr International Academy, University of Copenhagen \\
Blegdamsvej 17, 2100 K{\o}benhavn, Denmark}

\emailAdd{hjalte.frellesvig@nbi.ku.dk}

\abstract{In this paper we develop and demonstrate a method to obtain epsilon factorized differential equations for elliptic Feynman integrals. This method works by choosing an integral basis with the property that the period matrix obtained by integrating the basis over a complete set of integration cycles is diagonal. The method is a generalization of a similar method known to work for polylogarithmic Feynman integrals. We demonstrate the method explicitly for a number of Feynman integral families with an elliptic highest sector.}

\begin{document} 
\maketitle
\flushbottom

\section{Introduction}

The method of differential equations~\cite{Barucchi:1973zm,KOTIKOV1991158,Gehrmann:1999as} is the primary way phenomenologically relevant Feynman integrals are computed. This method works by taking the derivative with respect to a kinematic variable (or to several) of the members of a minimal basis of Feynman integrals for the problem, which we will refer to as master integrals. The result of taking that derivative may then be mapped back to the original master integrals using IBP identities~\cite{Chetyrkin:1981qh, Laporta:2001dd}, as implemented in for instance the computer codes FIRE~\cite{Smirnov:2008iw}, Kira~\cite{Maierhofer:2017gsa}, and others~\cite{Anastasiou:2004vj, vonManteuffel:2012np, Lee:2012cn}. This will result in a system of coupled linear differential equations for the master integrals, which may be solved with traditional methods. A breakthrough was made with the realization~\cite{Henn:2013pwa} that for a large class of Feynman integrals in dimensional regularization (with $d = d_0 - 2 \epsilon$), the solution of the system gets much simpler, in some cases trivialized, by choosing a special basis with the property that the differential equation system is epsilon factorized. Denoting the basis of master integrals as $J$, this means that the differential equation system in the kinematic variable $x$ may be written as
\begin{align}
\frac{\partial}{\partial x} J &= \epsilon A^{(x)} J
\label{eq:epsfac}
\end{align}
where $A^{(x)}$ is a matrix dependent on the kinematics but independent of $\epsilon$.

In the cases discussed in ref.~\cite{Henn:2013pwa}, these matrices have the additional property that the individual $A^{(x)}$-matrices may be unified as one matrix $A^{(x)} = \partial M/\partial x$ with the property that the entries of $M$ consist solely of logarithms of algebraic functions of the kinematics. A basis of master integrals for which the differential equations have these two properties (epsilon factorized, dlog form) is known as a canonical basis.

It is a well known fact that not all Feynman integrals can be brought to canonical form. One class (the simplest) of integrals for which a canonical form is unobtainable are those referred to as elliptic, a term that comes from the presence of elliptic curves at a cut surface. Such elliptic Feynman integrals, of which the fully massive sunrise integrals are a paradigmatic example, have been a subject of intense study in recent years~\cite{Broadhurst:1993mw,Berends:1993ee,Bauberger:1994nk,Bauberger:1994by,Bauberger:1994hx,Caffo:1998du,Laporta:2004rb,Groote:2005ay,Bailey:2008ib,MullerStach:2011ru,brown2011multiple,CaronHuot:2012ab,Groote:2012pa,Adams:2013nia,Bloch:2013tra,Remiddi:2013joa,Adams:2014vja,Broedel:2014vla,Bloch:2014qca,Adams:2015gva,Broedel:2015hia,Adams:2015ydq,Bloch:2016izu,Passarino:2016zcd,Primo:2016ebd,Remiddi:2016gno,Bonciani:2016qxi,Broadhurst:2016myo,Adams:2017ejb,vonManteuffel:2017hms,Ablinger:2017bjx,Remiddi:2017har,Hidding:2017jkk,Bourjaily:2017bsb,Broedel:2017kkb,Broedel:2017siw,Broedel:2017jdo,Lee:2017qql,Bogner:2017vim,Groote:2018rpb,Adams:2018bsn,Adams:2018kez,Adams:2018yfj,Ablinger:2018zwz,Broedel:2018qkq,Bogner:2019lfa,Bourjaily:2020hjv,Frellesvig:2021vdl,Bourjaily:2021vyj}. The purpose of this paper is to extend the epsilon factorized property of the differential equations given by eq.~\eqref{eq:epsfac} to such elliptic cases. This has been achieved in the literature for a few specific examples~\cite{Adams:2018yfj, Bogner:2019lfa} already, but this paper will present a more general algorithm for how to obtain such a basis.

For integrals for which a canonical form is obtainable, several approaches have been developed to finding it~\cite{Henn:2014qga,Lee:2014ioa,Argeri:2014qva,Gehrmann:2014bfa,Hoschele:2014qsa,Meyer:2016slj,Frellesvig:2017aai,WasserMSc,Prausa:2017ltv,Gituliar:2017vzm,Dlapa:2020cwj,Henn:2020lye,Chen:2020uyk,Dlapa:2021qsl}. One popular approach~\cite{Henn:2014qga,Frellesvig:2017aai,WasserMSc} includes the requirement that the master integrals are \textit{pure}~\cite{Arkani-Hamed:2010pyv}, which means that the value of the integral on each maximal cut (here in the sense of combinations of unitarity cuts that fix all degrees of freedom) is just a number, as opposed to a function of the kinematic variables. For elliptic Feynman integrals performing a maximal cut in that sense is impossible due to the presence of the elliptic curve. Yet a natural generalization of that cutting operation is an integral over a cycle, and such integrals can be performed also for the non-zero genus surfaces appearing for elliptic Feynman integrals (and beyond).

The algorithm proposed in this paper is the natural generalization of that maximal cut procedure, or more specifically it is to require that each master integral is non-vanishing on one and only one of the (members of a set of) basic cycles. This is similar to the prescriptive unitarity approach proposed in refs.~\cite{Bourjaily:2017wjl, Bourjaily:2021vyj} but here performed at the integral level for the fully dimensionally regulated integrals.

The use of $d$-dimensional unitarity cuts to investigate Feynman integrals and simplify their relations, has a long history. This was first done in the context of dimension shift relations~\cite{Lee:2012te}, and later extended to differential equations~\cite{Primo:2016ebd}, IBPs~\cite{Bosma:2017ens}, and more~\cite{Frellesvig:2017aai, Primo:2017ipr, Harley:2017qut}. In some of those works~\cite{Bosma:2017ens, Primo:2017ipr} the notion of using a diagonal period matrix as a condition for obtaining an epsilon factorized differential equation, a notion central to this work, was prefigured, and so was the suggestion to keep certain epsilon-dependent prefactors before doing the cut analysis~\cite{Frellesvig:2017aai}. Thus this paper may rightfully be regarded as the newest entry that series of works.

The dual vector space nature of the set of integrands $\varphi_i$ and integration contours $\gamma_j$~\cite{Smirnov:2010hn, Lee:2013hzt, Bitoun:2017nre, Bosma:2017ens}, was clarified in recent work on the connection between Feynman integrals and the mathematical discipline of intersection theory~\cite{Mastrolia:2018uzb, Frellesvig:2019kgj, Frellesvig:2019uqt, Frellesvig:2020qot, Mizera:2019vvs, Weinzierl:2020xyy, Caron-Huot:2021xqj, Caron-Huot:2021iev}, where the contours and integrands are viewed as representatives of a twisted de Rahm homology and cohomology respectively. This change of perspective makes it natural to look at the \textit{period matrix} $P_{ij} \!\sim\! \int_{\gamma_j} \!\! \varphi_i$ of pairings between integrands and contours, and the algorithm proposed here is naturally formulated in terms of that object.

After introducing the notation in higher detail, we will in sec.~\ref{sec:motivation} do two examples of the algorithm used in non-elliptic cases, while reformulating it in a way that generalizes to the elliptic case. Then in sec.~\ref{sec:algorithm} we will make the complete formulation of our algorithm, and in sec.~\ref{sec:examples} do a number of successful examples of its use for which many of the resulting expressions are added as an ancillary file. Finally in sec.~\ref{sec:discussion} we will discuss some loose ends and open questions, and summarize and conclude. In appendix~\ref{app:nptagain} we will redo one of the examples with a different approach.

\subsection{Notation and conventions}
\label{sec:notation}

We will in this work encounter the complete elliptic integrals of respectively first, second and third kind. They are defined as
\begin{align}
K(k^2) &= \int_{0}^{1} \! \frac{\id x}{\sqrt{(1-x^2)(1-k^2 x^2)}}\\
E(k^2) &= \int_{0}^{1} \! \frac{\sqrt{1-k^2 x^2} \; \id x}{\sqrt{1-x^2}} \\
\Pi(n^2,k^2) &= \int_{0}^{1} \! \frac{\id x}{(1-n^2 x^2)\sqrt{(1-x^2)(1-k^2 x^2)}}
\end{align}
where we use the Mathematica convention in which the arguments are $k^2$ and $n^2$.
The three complete elliptic integrals have the property that any integral of the form $\int_{\mathcal{C}} \mathcal{R}(x,Y) \id x$ can be written in terms of them (along with elementary functions), where $\mathcal{R}$ is any rational function, $Y$ is the square root of a polynomial in $x$ of degree three or four, and where $\mathcal{C}$ is any closed contour.

We will in this work only discuss two-loop Feynman integrals, which can be written as
\begin{align}
I_{\{a\}} &= \frac{1}{(2 \pi)^{d}} \int \frac{\id^d k_1 \id^d k_2}{\prod_i D_i^{a_i}}
\end{align}
where the propagators $D_i$ are quadratic functions of the momenta on the form $k^2{-}m^2$ and $d$ is the spacetime dimension. As mentioned earlier, we will consider the integrals in dimensional regularization where $d = d_0 - 2 \epsilon$ with $d_0$ being an integer, in this work either $2$ or $4$.

We will consider our Feynman integrals in Baikov representation, in either its standard~\cite{Baikov:1996iu, Lee:2010wea, Grozin:2011mt} or loop-by-loop~\cite{Frellesvig:2017aai, Frellesvig:2021vem} version. Thus the integral may be written
\begin{align}
I_{\{a\}} &= \mathcal{K} \int_{\mathcal{C}} \frac{u \, \id^n x}{\prod_i x_i^{a_i}}
\label{eq:Baikovdef}
\end{align}
where the $x_i$ are Baikov variables (that equal the original propagators, $x_i=D_i$), the $\alpha_j$ are generally irrational powers, and where $u$ is a multivalued function that will have the general form
\begin{align}
u = \prod_j \mathcal{B}_{j}^{\alpha_j}
\end{align}
\vspace{-4mm}

\noindent where the $\mathcal{B}_j$ are polynomial functions called Baikov polynomials of the $x_i$. The prefactor $\mathcal{K}$ will generally be given in terms of gamma functions of arguments linear in the spacetime dimension. In the examples we will look at, $\mathcal{K}$ will be a \textit{pure} function, by which we will mean a function with the property that the coefficients in the $\epsilon$ expansion has uniform \textit{weight}~\cite{Duhr:2014woa} such that the coefficient of $\eps^n$ has weight one higher than that of the coefficient of $\eps^{n-1}$ and only numerical prefactors. We will not make any attempts at generalizing these notions to elliptic objects.

We will in this paper make significant use of generalized unitarity cuts, in particular \textit{maximal cuts} which refers to cuts of all genuine propagators (and thus differs from the notation in the $\mathcal{N}{=}4$ literature, in which a maximal is defined to fix all degrees of freedom). This is due to the fact that differential equations~\cite{Primo:2016ebd}, as well as IBP relations~\cite{Bosma:2017ens}, remain the same after such a cutting procedure, and indeed the Baikov parametrization is particularly suited for generalized unitarity cuts at the integral level~\cite{Frellesvig:2017aai, Harley:2017qut}.
\nocite{Bosma:2017hrk}

In the examples in this paper, the maximal cut will leave a single integration to be done, bringing the integrals to a univariate form of eq.~\eqref{eq:Baikovdef}
\begin{align}
I_{\{a\}} |_{\text{max cut}} \ &= \mathcal{K} \int_{\mathcal{C}} u \, \hat{\phi} \, \id z
\end{align}
where $z$ is the remaining Baikov variable, and $\hat{\phi}$ is a rational function of $z$ and the kinematical variables. We will occasionally use the abbreviation $\phi := \hat{\phi} \id z$. In general this notation follows that of the recent work on intersection theory and Feynman integrals such as refs.~\cite{Mastrolia:2018uzb, Frellesvig:2019kgj}.

We will also discuss various integration contours. We will name them with the symbols $\mathcal{C}$ or $\gamma$, and use the notation $\mathcal{C}_i$ for a contour surrounding a pole at $z=i$, and the notation $\mathcal{C}_{\text{i-j}}$ for a contour surrounding a branch cut between the points $z=i$ and $z=j$.

\subsection{Notation - The sunrise integrals}
\label{sec:sunrisedef}

We will in this paper work quite a lot with the sunrise integral with different mass distributions. Rather than repeating the same definitions over and over, let us write them here in the most general case:

The three-mass sunrise integral is defined by the complete set of propagators
\begin{align}
D_1 &= k_2^2 - m_1^2 \,, & D_2 &= (k_1 - k_2)^2 - m_2^2 \,, & D_3 &= (k_1 + p)^2 - m_3^2 \,, \nonumber \\
D_4 &= (k_1)^2 \,, & D_5 &= (k_2 + p)^2 \,,
\label{eq:sunrisedef1}
\end{align}
with $p^2=s$, and where only the first three may appear as genuine propagators.

With these definition, the integrals are given as
\begin{align}
I^{\text{sunrise}}_{a_1a_2a_3a_4a_5} &= \int \frac{\id^d k_1 \id^d k_2}{(2 \pi)^{d}} \frac{D_4^{-a_4} D_5^{-a_5}}{D_1^{a_1} D_2^{a_2} D_3^{a_3}}
\label{eq:sunrisedef2}
\end{align}
We will always look at the case where $d = 2 - 2 \epsilon$, when discussing the sunrise.

\begin{figure}
\centering
\includegraphics[scale=0.90]{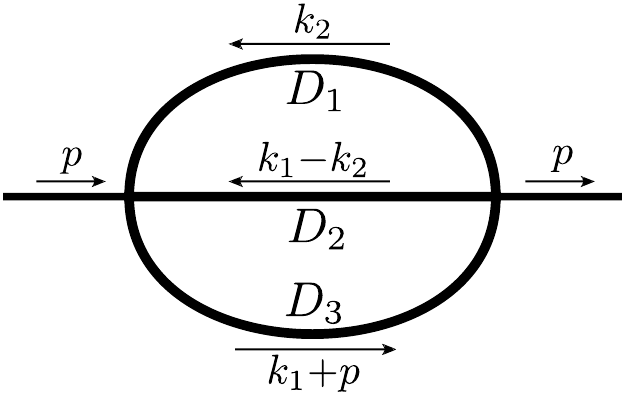}
\caption{The sunrise integrals}
\label{fig:sunrise}
\end{figure}

In the loop-by-loop Baikov representation, this integral may be written as
\begin{align}
I^{\text{sunrise}}_{a_1a_2a_3a_40} &= \mathcal{K} \int_{\mathcal{C}} \frac{x_4^{-a_4} \, u \, \id^4 x}{x_1^{a_1} x_2^{a_2} x_3^{a_3}}
\label{eq:sunrisedef3}
\end{align}
where
\begin{align}
u &= x_4^{\epsilon} \times \big( (m^2_1 {-} m^2_2 {+} x_1 {-} x_2)^2 {-} 2 (m^2_1 {+} m^2_2 {+} x_1 {+} x_2) x_4 {+} x_4^2 \big)^{-\frac12-\epsilon} \nonumber \\
& \;\;\;\; \times \big( (m^2_3 {-} s {+} x_3)^2 {-} 2 (m^2_3 {+} s {+} x_3) x_4 {+} x_4^2 \big)^{-\frac12-\epsilon}
\label{eq:sunriseu}
\end{align}
is the product of Baikov polynomials raised to their corresponding powers, and where
\begin{align}
\mathcal{K} &= \frac{2^{4 \epsilon}}{\pi \Gamma^2 (\tfrac12 - \epsilon)}
\label{eq:Ksunrise}
\end{align}
is a pure function. The choice $a_5=0$ made in eq.~\eqref{eq:sunrisedef3} is done in order to simplify the loop-by-loop parametrization, which in this case is performed by first integrating out the $k_2$-loop. This limitation of the loop-by-loop Baikov parametrization can be overcome (for instance using a Passarino-Veltman~\cite{Passarino:1978jh} inspired approach as described in section 11.1 of ref.~\cite{Frellesvig:2019kgj}, and for a recent alternative approach to this see ref.~\cite{Chen:2022lzr},) but that additional complication is not necessary for the present discussion.

On the maximal cut (i.e. of $D_1$, $D_2$, $D_3$, and with $z=D_4$), $u$ reduces to
\begin{align}
u|_{3 \times \text{cut}} &= z^{\epsilon} \; \big( (m^2_1 {-} m^2_2)^2 {-} 2 (m^2_1 {+} m^2_2) z {+} z^2 \big)^{-\frac12-\epsilon} \; \big( (m^2_3 {-} s)^2 {-} 2 (m^2_3 {+} s) z {+} z^2 \big)^{-\frac12-\epsilon}
\label{eq:sunriseucut}
\end{align}
meaning that any integral in the sunrise family can be written as
\begin{align}
I^{\text{sunrise}} &= \mathcal{K} \int_{\mathcal{C}} u|_{3 \times \text{cut}} \, \hat{\phi} \, \id z
\label{eq:sunrisedef6}
\end{align}
on that cut, where $\hat{\phi}$ will be some rational function of $z$, the exact form of which will depend on the values of the propagator powers $a_i$.

We note that $u|_{3 \times \text{cut}}$ is a pure function divided by the square root of a polynomial of degree four, a fact we will be using implicitly in the following.

\section{Motivation}
\label{sec:motivation}

In this section we will discuss two simple examples of non-elliptic Feynman integrals, and show how to find their canonical forms with the traditional method. This method will then be reformulated in terms of diagonalization of the period matrix, giving a formulation that generalizes directly to the elliptic case.

\subsection{The double box}

Let us as an illustration of the procedure discuss the well-known fully massless double box. That family of Feynman integrals (first computed in~\cite{Smirnov:1999gc, Smirnov:1999wz}) was used as the example in ref. \cite{Henn:2013pwa} which introduced the concept of epsilon-factorized differential equations. It is given by the set of propagators
\begin{align}
D_1 &= k_1^2\,, & D_2 &= (k_1{+}p_1)^2\,, & D_3 &= (k_1{+}p_1{+}p_2)^2\,, & D_4 &= (k_2{+}p_1{+}p_2)^2\,, & D_5 &= (k_2{-}p_4)^2\,, \nonumber \\
D_6 &= k_2^2\,, & D_7 &= (k_1{-}k_2)^2\,, & D_8 &= (k_1{-}p_4)^2\,, & D_9 &= (k_2{+}p_1)^2\,, &
\end{align}
of which only the first seven are allowed to appear as genuine propagators.

The family contains eight master integrals of which two are in the highest sector, and it is those two that will interest us in the following. In ref.~\cite{Henn:2013pwa} it was found that a set of integrals giving epsilon-factorized differential equations are
\begin{align}
J_1 = c s^2 t I^{\text{db}}_{111111100} \qquad \text{and} \qquad J_2 = c s^2 I^{\text{db}}_{1111111 \sminus 10} 
\label{eq:dbhenn}
\end{align}
where $c = \eps^4 s^{-2 \eps}$ is a pure prefactor to be ignored in the following.
How can we know \textit{a priori} the two prefactors $s^2 t$ and $s^2$?

One way is to perform the maximal cut (in the sense of cutting the seven actual propagators) of the integrals in the family. Using the loop-by-loop Baikov parametrization, this gives
\begin{align}
I^{\text{db}}_{1111111a0}|_{\text{7-cut}} &= \frac{-\mathcal{K}}{s^2} \int (z-t)^{-1 - 2 \epsilon} z^{-a-1-\epsilon} (s + z)^{\epsilon} \, \id z
\end{align}
where
\begin{align}
\mathcal{K} &= \frac{4^{\sminus 1+2 \eps} s^{\sminus 2 \eps} \, t^{\eps} \, (s{+}t)^{\eps}}{\pi^3 \Gamma^2( \tfrac{1}{2} {-} \eps )}
\end{align}
is a pure function.

The prescription is now to factorize out the pure part of the integrand, which here corresponds to taking the limit $\epsilon \rightarrow 0$ of the object under the integral sign. This (ignoring the already pure $\mathcal{K}$) gives the integrand
\begin{align}
\hat{\Phi} &= \frac{-z^{-a}}{s^2 z (z-t)}
\end{align}
Continuing the cutting procedure shows us that $I^{\text{db}}_{111111100}$ has two residues. One at $z=0$ evaluating to $1/(s^2t)$ and the other at $z=t$ evaluating to $-1/(s^2t)$. On the other hand $I^{\text{db}}_{1111111 \sminus 10}$ also has two residues, one at $z=t$ evaluating to $-1/(s^2)$ and the other at $z=\infty$ evaluating to $1/(s^2)$. So we see that the two integrals become pure if $I^{\text{db}}_{111111100}$ is given a prefactor of $s^2t$ and $I^{\text{db}}_{1111111 \sminus 10}$ a prefactor of $s^2$, as it is given by eq.~\eqref{eq:dbhenn}.

\begin{figure}
\centering
\includegraphics[scale=1.00]{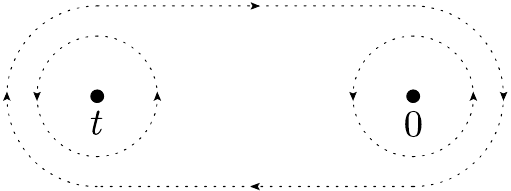}
\caption{Contours discussed for the double box}
\label{fig:condb}
\end{figure}

A different way of formulating the procedure is by writing up the period matrix for the two integrals, again after factorizing out and discarding the pure part of $u$. We pick the two integration cycles as $\gamma_1$ being the cycle $\mathcal{C}_0$ surrounding the $(z{=}0)$-pole and $\gamma_2$ being $\mathcal{C}_{\infty}$ surrounding the $(z{=}\infty)$-pole, with the various contours shown on fig.~\ref{fig:condb}. Giving the two integrals prefactors $f_1$ and $f_2$ we get the period matrix
\begin{align}
P_{ij} = \int_{\gamma_j} \!\! f_i \hat{\Phi}_i \id z \quad \Rightarrow \quad P = 2 \pi i \left[ \begin{array}{cc} \frac{f_1}{s^2 t} & 0 \\ 0 & \frac{f_2}{s^2} \end{array} \right] 
\end{align}
and requiring the period matrix to be $2 \pi i$ times the unit matrix $I$ fixes the $f_i$ to the values found above.

Making this procedure more general and algorithmic, we could also write each of the canonical integrals we want to find as general linear combinations of the two precanonical intermediate basis integrals $I^{\text{db}}_{111111100}$ and $I^{\text{db}}_{1111111 \sminus 10}$, that is
\begin{align}
J_1 = f_{11} I^{\text{db}}_{111111100} + f_{12} I^{\text{db}}_{1111111 \sminus 10}  \nonumber \\
J_2 = f_{21} I^{\text{db}}_{111111100} + f_{22} I^{\text{db}}_{1111111 \sminus 10}
\end{align}
This gives the period matrix
\begin{align}
P = 2 \pi i \left[ \begin{array}{cc} \frac{f_{11}}{s^2 t} & \frac{f_{12}}{s^2} \\[1mm] \frac{f_{21}}{s^2 t} & \frac{f_{22}}{s^2} \end{array} \right] 
\end{align}
and once again imposing $P = 2 \pi i I$ fixes the $f_{ij}$ coefficients uniquely to
\begin{align}
f_{11} = s^2t \,,\qquad f_{12} = 0 \,,\qquad f_{21} = 0 \,,\qquad f_{22} = s^2 \,,
\end{align}
again corresponding to the prefactors given by eq.~\eqref{eq:dbhenn}.

\subsection{The two-mass non-elliptic sunrise}

Let us look at another non-elliptic example that looks more similar to the elliptic examples we will encounter later.
That is the two-mass non-elliptic sunrise integral (sne), defined as in eqs. \eqref{eq:sunrisedef1} and \eqref{eq:sunrisedef2} but with $m_3^2 = 0$ and $m_1^2 = m_2^2 \define m^2$.

\begin{figure}
\centering
\includegraphics[scale=1.00]{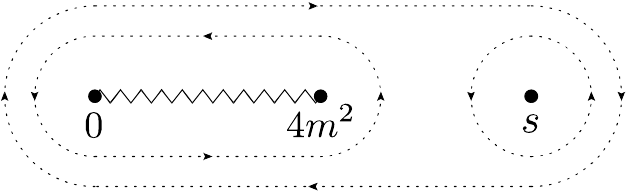}
\caption{Contours discussed for the two-mass non-elliptic sunrise}
\label{fig:consne}
\end{figure}

Using the loop-by-loop Baikov parametrization of this integral as in eq.~\eqref{eq:sunrisedef3}, we may make a univariate representation of the integral on the maximal cut. Following our prescription we have to factorize out the pure part from the integrand in order to analyze its behaviour. The integrand then becomes (from eq.~\eqref{eq:sunrisedef6})
\begin{align}
U \, \hat{\phi} \qquad \text{with} \qquad U = \frac{1}{(z-s) \sqrt{z (z - 4 m^2)}}
\end{align}
If one were to follow the prescription of cutting the last variable in order to impose a maximum cut of $\pm 1$, one would (formally) first have to get rid of the square root through rationalization. This can be done by a variable change such as
\begin{align}
z \; \rightarrow \; m^2 \frac{(1+y)^2}{y}
\end{align}
but let us try to proceed without such procedures.
The integration plane can be seen depicted on fig.~\ref{fig:consne}. One can define a basis of independent cycles as a cycle $\gamma_1$ surrounding the $(z=s)$-pole, and one $\gamma_2$ surrounding the pole at infinity. Picking the contour around the branch-cut from $z=0$ to $z = 4 m^2$ as one of the master contours would be equally valid, but the above choice gives the nicest result.

We will make a choice of intermediate basis integrals (precanonicals) as $I^{\text{sne}}_{11100}$ and $I^{\text{sne}}_{111 \sminus 10}$. This corresponds to the integrands
\begin{align}
\hat{\phi}_1 = 1 \qquad \text{and} \qquad \hat{\phi}_2 = z
\end{align}
With this we may compute the integrals on the cycles, giving
\begin{align}
g_{11} &= \int_{\gamma_1} \! U \hat{\phi}_1 \id z = \frac{2 \pi i}{\sqrt{s (s - 4 m^2)}}    & g_{12} &= \int_{\gamma_2} \! U \hat{\phi}_1 \id z = 0 \nonumber \\
g_{21} &= \int_{\gamma_1} \! U \hat{\phi}_2 \id z = \frac{2 \pi i s}{\sqrt{s (s - 4 m^2)}}  & g_{22} &= \int_{\gamma_2} \! U \hat{\phi}_2 \id z = -2 \pi i
\end{align}
Writing the two integrals we are looking for as generic linear combinations
\begin{align}
J_1 = f_{11} I^{\text{sne}}_{11100} + f_{12} I^{\text{sne}}_{111 \sminus 10} \qquad\quad \text{and} \qquad\quad J_2 = f_{21} I^{\text{sne}}_{11100} + f_{22} I^{\text{sne}}_{111 \sminus 10}
\end{align}
we obtain the period matrix $P_{ij} = f_{ik} g_{kj}$ as
\begin{align}
P &= \left[ \begin{array}{cc}
\frac{2 \pi i ( f_{11} + s f_{12} )}{\sqrt{s (s - 4 m^2)}}\;\; & -2 \pi i f_{12} \\ \frac{2 \pi i ( f_{21} + s f_{22} )}{\sqrt{s (s - 4 m^2)}}\;\; & -2 \pi i f_{22} \end{array} \right]
\end{align}
Solving for $P=2\pi i I$ gives
\begin{align}
f_{11} = \sqrt{s (s {-} 4 m^2)} \,,\qquad f_{12} = 0 \,,\qquad f_{21} = s \,,\qquad f_{22} = -1 \,,
\end{align}
corresponding to
\begin{align}
J_1 = \sqrt{s(s {-} 4 m^2)} I^{\text{sne}}_{11100} \qquad \text{and} \qquad J_2 = s I^{\text{sne}}_{11100} - I^{\text{sne}}_{111 \sminus 10}
\end{align}
With this we may compute the epsilon factorized differential equation matrix
\begin{align}
\frac{\id J_i}{\id s} = \epsilon A_{ij} J_j \qquad \text{with} \qquad A = \left[ \begin{array}{cc}
\frac{4 (s {-} m^2)}{(4 m^2 {-} s) s} & \frac{3}{\sqrt{s (s {-} 4 m^2)}} \\ \frac{-2}{\sqrt{s (s {-} 4 m^2)}} & \frac{1}{s} \end{array} \right]
\end{align}
written disregarding integrals in lower sectors (here only the double-tadpole $I^{\text{sne}}_{11000}$) so we see that our method works in this case as well. This success is what motivates us to try the same approach in elliptic cases.

\section{The proposed algorithm}
\label{sec:algorithm}

Motivated by the examples in the previous section, we are now ready to formulate our algorithm. Having a set of $\nu$ integrals of the form
\begin{align}
J_i &= \mathcal{K} \int_{\mathcal{C}} u \hat{\varphi}_i \id^n x
\end{align}
where as discussed in section~\ref{sec:notation} $u$ is a multivalued function, and $\hat{\varphi}_i$ a set of $\nu$ rational (in $x$) functions, our claim is that if the integrand can be written as
\begin{align}
\mathcal{K} u \hat{\varphi}_i &= \sigma \hat{\Phi}_i
\end{align}
where $\sigma$ is a pure function, and where the period matrix $P$ with $P_{ij} = \int_{\gamma_j} \!\! \hat{\Phi}_i \id^n x$ is proportional to the identity matrix as $P = (2 \pi i)^n I$, then the integrals $J_i$ will fulfill epsilon-factorized differential equations. In cases where $\mathcal{K}$ is itself pure to begin with, it can be ignored in the above discussion.

What we will do in practice is to write the basis integrals as a linear combination of known intermediate basis integrals $I_j$ (often called precanonical in the polylogarithmic case,) that is $J_i = f_{ij} I_j$ or correspondingly as $\hat{\Phi}_i = f_{ij} \hat{\Phi}^{\text{int}}_j$, and then imposing $P = (2 \pi i)^n I$ will impose $\nu^2$ constraints on the $f_{ij}$ fixing them all uniquely.

In the examples in the following section, we will write the $\hat{\Phi}^{\text{int}}_{\;}$ as $\hat{\Phi}^{\text{int}}_i = \hat{\phi}_i/Y$ where $Y$ is the square root of a (monic) polynomial of degree four, corresponding to the elliptic curve characterizing the problem.

\section{Examples}
\label{sec:examples}

In this section we will look at examples of elliptic Feynman integrals which we can bring into a form that has epsilon-factorized differential equations with the above algorithm. We will restrict the discussion to cases where the ellipticity is present in the highest sector only, and where a maximal cut can bring the integrals in that sector to a univariate form.

\subsection{The same-mass elliptic sunrise}
\label{sec:samemass}

As a first example let us look at the most basic of elliptic Feynman integrals, the same-mass elliptic sunrise (s1m). This is defined as in section~\ref{sec:sunrisedef}, but with the restriction that the three masses are the same, i.e. $m_1^2=m_2^2=m_3^2=m^2$.

This integral family has three master integrals. One is the double-tadpole (e.g. $I^{\text{s1m}}_{11000}$) while the last two are in the highest sector and the only ones we will care about in the following where we disregard subsectors completely.
Picking as intermediate basis integrals $I^{\text{s1m}}_{11100}$ and $I^{\text{s1m}}_{21100}$ we can write the two candidate integrals as
\begin{align}
J_1 &= c (f_{11} I^{\text{s1m}}_{11100} + f_{12} I^{\text{s1m}}_{21100}) \, + \, \text{lower} \nonumber \\
J_2 &= c (f_{21} I^{\text{s1m}}_{11100} + f_{22} I^{\text{s1m}}_{21100}) \, + \, \text{lower} \label{eq:s1mdef}
\end{align}
where the ``lower'' covers potential contributions from the double-tadpole sector. The $c$ is a constant prefactor, the value of which will not change the differential equations and which will be ignored in all of the following.

Our goal now is to find values of the $f_{ij}$ that makes the period matrix of these two integrals the identity matrix.

Parametrizing this integral with the loop-by-loop Baikov parametrization, we get from eq.~\eqref{eq:sunrisedef3}
\begin{align}
I^{\text{s1m}}_{a_1a_2a_3a_40} \; &= \mathcal{K} \int_{\mathcal{C}} \id^4 x \frac{x_4^{-a_4} \; u}{x_1^{a_1} x_2^{a_2} x_3^{a_3}}
\end{align}
with
\begin{align}
u &= x_4^{\epsilon} \times \big( (x_1 {-} x_2)^2 {-} 2 (2 m^2 {+} x_1 {+} x_2) x_4 {+} x_4^2 \big)^{-\frac12-\epsilon} \nonumber \\
& \;\;\;\; \times \big( (m^2 {-} s {+} x_3)^2 {-} 2 (m^2 {+} s {+} x_3) x_4 {+} x_4^2 \big)^{-\frac12-\epsilon}
\end{align}
as by eq.~\eqref{eq:sunriseu}. $\mathcal{K}$ given by eq.~\eqref{eq:Ksunrise} is a pure function, and $u$ is pure multiplied with a factor of $Y^{-1}$ where on the maximal cut (i.e. of $x_1$,$x_2$,$x_3$ and with $x_4 \rightarrow z$) 
\begin{align}
Y &= \sqrt{z \, (z{-}4m^2) \, \big(z^2 - 2 (m^2{+}s) z + (m^2{-}s)^2 \big)} \,,
\end{align}
Following the algorithm of sec.~\ref{sec:algorithm} we factor out the pure part of $u$, leaving the expressions
\begin{align}
I^{\text{s1m}}_{11100} \rightarrow \int_{\mathcal{C}} \frac{\hat{\phi}_1 \id z}{Y} \qquad\qquad\qquad I^{\text{s1m}}_{21100} \rightarrow \int_{\mathcal{C}} \frac{\hat{\phi}_2 \id z}{Y}
\end{align}
where the integrands are given by
\begin{align}
\hat{\phi}_1 = 1 \qquad \text{and} \qquad \hat{\phi}_2 = \frac{1 {+} 2 \epsilon}{z {-} 4 m^2}
\end{align}
of which the latter has been computed as $\hat{\phi}_2 = (\partial_{x_1} u)/u |_{\text{cut}}$.

Please note the presence of the $\epsilon$-dependence in $\hat{\phi}_2$. This factor comes out of the algorithm in a natural way, and is essential in order to get the correct epsilon factorization of the differential equation. Such $\epsilon$-dependence would be missed by a purely two-dimensional approach!

\begin{figure}
\centering
\includegraphics[scale=1.00]{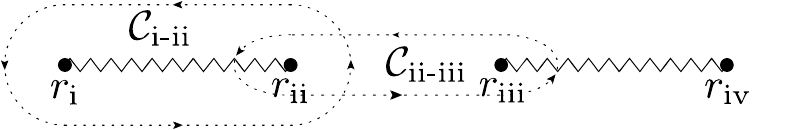}
\caption{The two independent contours for the same-mass elliptic sunrise}
\label{fig:cons1m}
\end{figure}

$Y^2$ has four roots
\begin{align}
r_{\text{i}} &= 0 \;, & r_{\text{ii}} &= (\sqrt{m^2} - \sqrt{s})^2 \;, & r_{\text{iii}} &= (\sqrt{m^2} + \sqrt{s})^2 \;, & r_{\text{iv}} &= 4 m^2 \;,
\end{align}
so we can look at the two basic cycles $\gamma_1 = \mathcal{C}_{\text{ii-iii}}$ and $\gamma_2 = \mathcal{C}_{\text{i-ii}}$ which may be seen on fig.~\ref{fig:cons1m}. In the convergent case
\begin{align}
\int_{\mathcal{C}_{i\text{-}j}} &= \;\; 2 \int_{r_i}^{r_j}
\end{align}
allowing us to compute the integrals
\begin{align}
g_{11} &= \int_{\gamma_1} \frac{\hat{\phi}_1 \id z}{Y} \!\! &&= \, \frac{4 K (k^2)}{R} \nonumber \\
g_{21} &= \int_{\gamma_1} \frac{\hat{\phi}_2 \id z}{Y} \!\! &&= \, \frac{(1{+}2 \epsilon)}{m^2 R} \left( K(k^2) + \frac{(\sqrt{m^2}{+}\sqrt{s})^2}{4 m^2 - (\sqrt{m^2}{+}\sqrt{s})^2} E(k^2) \right) \nonumber \\
g_{12} &= \int_{\gamma_2} \frac{\hat{\phi}_1 \id z}{Y} \!\! &&= \, \frac{-4i K (1{-}k^2)}{R} \\
g_{22} &= \int_{\gamma_2} \frac{\hat{\phi}_2 \id z}{Y} \!\! &&= \, \frac{-2i (1{+}2 \epsilon)}{R \big( 4 m^2 {-} (\sqrt{m^2}{+}\sqrt{s})^2 \big) } \left( 2 K (1{-}k^2) - \frac{(\sqrt{m^2}{+}\sqrt{s})^2}{2 m^2} E (1{-}k^2) \right) \nonumber
\end{align}
with
\begin{align}
R &= \sqrt{(3 \sqrt{m^2}{-}\sqrt{s})(\sqrt{m^2}{+}\sqrt{s})^3}
\end{align}
and
\begin{align}
k^2 &:= \frac{16 \sqrt{ (m^2)^3 s }}{(3 \sqrt{m^2}{-}\sqrt{s})(\sqrt{m^2}{+}\sqrt{s})^3} & 1{-}k^2 &= \frac{(3 \sqrt{m^2}{+}\sqrt{s})(\sqrt{m^2}{-}\sqrt{s})^3}{(3 \sqrt{m^2}{-}\sqrt{s})(\sqrt{m^2}{+}\sqrt{s})^3}
\end{align}

The period matrix ($P_{ij} = \int_{\gamma_j} \phi_i/Y$) of the two integrals of eqs.~\eqref{eq:s1mdef} is then

\begin{align}
P &= \left[ \begin{array}{cc} f_{11} g_{11} + f_{12} g_{21} \;\;\; & f_{11} g_{12} + f_{12} g_{22}  \\[2mm]
f_{21} g_{11} + f_{22} g_{21} \;\;\; & f_{21} g_{12} + f_{22} g_{22}  \end{array} \right]
\end{align}
Imposing $M=2 \pi i I$ gives a unique solution for the $f_{ij}$ as
\begin{align}
f_{11} &= i R \, E(1{-}k^2) - \frac{4 i m^2 R}{(\sqrt{m^2}{+}\sqrt{s})^2} K(1{-}k^2) \nonumber \\
f_{12} &= \frac{-4i m^2 (\sqrt{m^2}{-}\sqrt{s}) (3 \sqrt{m^2}{+}\sqrt{s}) R}{(1{+}2 \epsilon) (\sqrt{m^2}{+}\sqrt{s})^2} K(1{-}k^2) \nonumber \\
f_{21} &= -R \, E(k^2) - \frac{(\sqrt{m^2}{-}\sqrt{s}) (3 \sqrt{m^2}{+}\sqrt{s}) R}{(\sqrt{m^2}{+}\sqrt{s})^2} K(k^2) \label{eq:fsols1m} \\
f_{22} &= \frac{-4 m^2 (\sqrt{m^2}{-}\sqrt{s}) (3 \sqrt{m^2}{+}\sqrt{s}) R}{(1{+}2 \epsilon) (\sqrt{m^2}{+}\sqrt{s})^2} K(k^2) \nonumber
\end{align}

With this choice we can compute the system of differential equations for the two integrals. It is epsilon factorized
\begin{align}
\frac{\id J_i}{\id s} = \epsilon A_{ij} J_j \; + \; \text{lower}
\end{align}
with
\begin{align}
A_{11} &= \frac{-1}{2s} - \frac{(3 m^2{+}s)^2 K(k^2) K(1{-}k^2)}{\pi s (3 \sqrt{m^2}{-}\sqrt{s}) (\sqrt{m^2}{+}\sqrt{s})^3} + \frac{3 (\sqrt{m^2}{+}\sqrt{s})^2 \, E(k^2) E(1{-}k^2)}{\pi s (\sqrt{m^2}{-}\sqrt{s}) (3\sqrt{m^2}{+}\sqrt{s})} \nonumber \\
& \;\;\; + \frac{(27 m^2 - 24 m^3 \sqrt{s} - 18 m^2 s - s^2) E(k^2) K(1{-}k^2)}{\pi s (m^2 {-} s) (9 m^2 {-} s)} \nonumber
\end{align}
\vspace{-4mm}
\begin{align}
A_{12} &= \frac{-16 i m^2 (3 m^3 {-} 3 m^2 \sqrt{s} {-} 3 m s {-} \sqrt{s}^3) K(1{-}k^2)^2}{\pi \sqrt{s} (m^2{-}s) (3 m^2{-}s) (\sqrt{m^2}{-}\sqrt{s})^2} + \frac{3 i (\sqrt{m^2}{+}\sqrt{s})^2 E(1{-}k^2)^2}{\pi s (\sqrt{m^2}{-}\sqrt{s}) (3 \sqrt{m^2}{+}\sqrt{s})} \nonumber \\
& \;\;\; + \frac{2 i (3 m^3 - 9 m^2 \sqrt{s} - 2 ms + \sqrt{s}^3)  K(1{-}k^2) E(1{-}k^2)}{\pi s (m^2{-}s) (3 \sqrt{m^2}{-}\sqrt{s})} \nonumber \\
A_{21} &= \frac{i (3 m^2 {+} s)^2 K(k)^2}{\pi s (3 \sqrt{m^2}{-}\sqrt{s}) (\sqrt{m^2}{+}\sqrt{s})^3} + \frac{3 i (\sqrt{m^2}{+}\sqrt{s})^2 E(k)^2}{\pi s (3 \sqrt{m^2}{+}\sqrt{s}) (\sqrt{m^2}{-}\sqrt{s})} \nonumber \\
& \;\;\; - \frac{4 i (3m^2{-}s) (3 m^2{+}s) K(k^2) E(k^2)}{\pi s (m^2{-}s) (9m^2{-}s)} \label{eq:difeqs1m}
\end{align}
\vspace{-4mm}
\begin{align}
A_{22} &= \frac{2 (2 \sqrt{m^2} + \sqrt{s})}{\sqrt{s} (\sqrt{m^2}{-}\sqrt{s}) (3 \sqrt{m^2}{+}\sqrt{s})} - \frac{16 m^2 ( 3 m^3 {-} 3 m^2 \sqrt{s} {-} 3 m s {-} \sqrt{s}^3) K(k^2) K(1{-}k^2)}{\pi \sqrt{s} (m^2{-}s) (9 m^2{-}s) (\sqrt{m^2}{+}\sqrt{s})^2} \nonumber \\
& \;\;\; - \frac{3 (\sqrt{m^2}{+}\sqrt{s})^2 E(k^2) E(1{-}k^2)}{\pi s (\sqrt{m^2}{-}\sqrt{s}) (3 \sqrt{m^2}{+}\sqrt{s})} + \frac{(27 m^4 {-} 24 m^3 \sqrt{s} {-} 18 m^2 s {-} s^2) K(k^2) E(1{-}k^2)}{\pi s (m^2{-}s) (9 m^2 {-} s)} \nonumber
\end{align}
and a similar epsilon-factorized differential equation may be found in the other variable $\id J/ \id m^2$. This shows that our algorithm works for elliptic examples too.

\subsection{The elliptic nonplanar double triangle}
\label{sec:npt}

The elliptic nonplanar double triangle (npt)~\cite{Czakon:2008ii, vonManteuffel:2017hms}, is defined by the set of propagators
\begin{align}
D_1 &= k_1^2 {-} m^2, & D_2 &= (k_1{+}p_1)^2 {-} m^2, & D_3 &= (k_1{-}k_2{-}p_2)^2 {-} m^2, & D_4 &= (k_1{-}k_2)^2 {-} m^2, \nonumber \\
D_5 &= k_2^2, & D_6 &= (k_2{+}p_1{+}p_2)^2, & D_7 &= (k_2{+}p_2)^2,
\end{align}
where only the first six may appear as actual propagators.
The kinematics is such that $p_1^2=p_2^2 = 0$ and $(p_1{+}p_2)^2 = s$.

\begin{figure}
\centering
\includegraphics[scale=0.90]{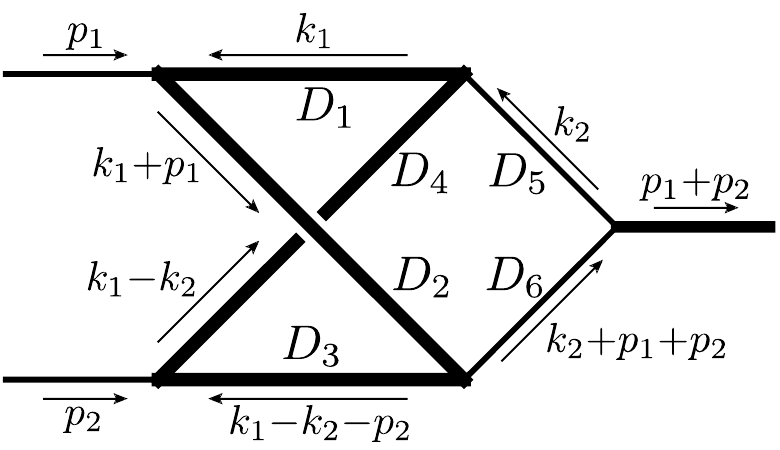}
\caption{The elliptic nonplanar double triangle}
\label{fig:npt}
\end{figure}

This sector contains 11 master integrals. The first nine are in subsectors, and can be put in canonical form in the traditional sense. The last two are in the highest, elliptic sector, and it is those that will concern us here. Choosing as intermediate basis integrals the set $I^{\text{npt}}_{1111110}$ and $I^{\text{npt}}_{2111110}$ we will write the two elliptic master integrals as linear combination of the intermediate basis:
\begin{align}
J_1 &= c (f_{11} I^{\text{npt}}_{1111110} + f_{12} I^{\text{npt}}_{2111110}) \, + \, \text{lower} \nonumber \\
J_2 &= c (f_{21} I^{\text{npt}}_{1111110} + f_{22} I^{\text{npt}}_{2111110}) \, + \, \text{lower} \label{eq:npedef}
\end{align}

Using the Baikov parametrization, the integrals in this sector may on the maximal cut (of $x_1$-$x_6$ with $z=x_7$) be written as
\begin{align}
I^{\text{npt}} \; &= \mathcal{K} \int_{\mathcal{C}} \! u \hat{\phi} \, \id z
\end{align}
Around $d=4$, we have once again that $\mathcal{K}$ is pure, and $u$ is pure times a factor of $Y^{-1}$ with
\begin{align}
Y &= \sqrt{z \, (z{+}s) \, (z^2 {+} sz {-} 4 m^2 s )} \,,
\end{align}
where a factor of $s$ has been absorbed in the $\hat{\phi}$. This justifies studying the integrals with the $u \rightarrow Y^{-1}$ limit taken, and in that case we may write the intermediate basis integrals as
\begin{align}
I^{\text{npt}}_{1111110} \rightarrow \int_{\mathcal{C}} \frac{\hat{\phi}_1 \id z}{Y} \qquad\qquad\qquad I^{\text{npt}}_{2111110} \rightarrow \int_{\mathcal{C}} \frac{\hat{\phi}_2 \id z}{Y}
\end{align}
where
\begin{align}
\hat{\phi}_1 = \frac{1}{s} \qquad \text{and} \qquad \hat{\phi}_2 = \frac{(1 {+} 2 \epsilon) (z {+} s)}{s (z^2 {+} sz {-} 4 m^2 s )}
\end{align}
The four roots of $Y^2$ are given as
\begin{align}
r_{\text{i}} = - \tfrac{1}{2} \sqrt{s} (\sqrt{s} {+} \sqrt{16 m^2 {+} s}) \,,\qquad r_{\text{ii}} = -s \,,\qquad r_{\text{iii}} = 0 \,,\qquad r_{\text{iv}} = - \tfrac{1}{2} \sqrt{s} (\sqrt{s} {-} \sqrt{16 m^2 {+} s}) \,,
\end{align}
and we see that a set of independent integration contours can be chosen as
\begin{align}
\gamma_1 &= \mathcal{C}_{\text{ii-iii}} & \gamma_2 &= \mathcal{C}_{\text{i-ii}}
\label{eq:gammanpt}
\end{align}
similarly to the previous example where the contours are depicted on fig.~\ref{fig:cons1m}.

We note that where in the previous section we always had $\int_{\mathcal{C}_{\alpha \text{-} \beta}} \phi = 2 \int_{\alpha}^{\beta} \phi$, that is not the case here for $\int_{\mathcal{C}_{\text{i-ii}}} \phi_2/Y$ due to a divergence in $z=r_{\text{i}}$. The results for the integrals are
\begin{align}
g_{11} &= \int_{\gamma_{1}} \!\! \frac{\hat{\phi}_1 \id z}{Y} \!\!\! &&=\;\; \frac{8 K (k^2)}{s^{3/2} ( \sqrt{16 m^2 {+} s} {+} \sqrt{s} ) } \nonumber \\
g_{21} &= \int_{\gamma_{1}} \!\! \frac{\hat{\phi}_2 \id z}{Y} \!\!\! &&=\;\; \frac{- 8 (1{+}2 \epsilon)}{s^{3/2} (16 m^2 {+} s) ( \sqrt{16 m^2 {+} s} {+} \sqrt{s} )} \left( K (k^2) + \frac{\sqrt{16 m^2 {+} s} {+} \sqrt{s}}{\sqrt{16 m^2 {+} s} {-} \sqrt{s}} \, E (k^2) \right) \nonumber \\
g_{12} &= \int_{\gamma_{2}} \!\! \frac{\hat{\phi}_1 \id z}{Y} \!\!\! &&=\;\; \frac{-8i K (1{-}k^2)}{s^{3/2} ( \sqrt{16 m^2 {+} s} {+} \sqrt{s} )} \\
g_{22} &= \int_{\gamma_{2}} \!\! \frac{\hat{\phi}_2 \id z}{Y} \!\!\! &&=\;\; \frac{i (1{+}2 \epsilon) }{s^{3/2} m^2 } \left( \frac{K (1{-}k^2)}{\sqrt{16 m^2 {+} s}} - \frac{\sqrt{16 m^2 {+} s} {+} \sqrt{s}}{2 (16 m^2 {+} s)} E (1{-}k^2) \right) \nonumber
\end{align}
where we have
\begin{align}
k^2 &:= \frac{4 \sqrt{s} \sqrt{16 m^2 {+} s}}{( \sqrt{16 m^2 {+} s} {+} \sqrt{s} )^2} \,, & 1{-}k^2 &= \frac{(\sqrt{16 m^2 {+} s} {-} \sqrt{s})^2}{( \sqrt{16 m^2 {+} s} {+} \sqrt{s} )^2} \,.
\end{align}

We may now compute the period matrix $P_{ij} = f_{ik} g_{kj}$, and imposing $P = 2 \pi i I$ allows us to uniquely fix the coefficients $f_{ik}$ of eq.~\eqref{eq:npedef} to
\begin{align}
f_{11} &= \tfrac{1}{2} i s^{3/2} (\sqrt{16 m^2{+}s} {+} \sqrt{s}) E(1{-}k^2) - i s^{3/2} \sqrt{16 m^2{+}s} K(1{-}k^2) \nonumber \\
f_{12} &= \frac{-i s^{3/2} (16 m^2 {+} s) (\sqrt{16 m^2 {+} s} {-} \sqrt{s}) K(1{-}k^2)}{2 (1{+}2 \epsilon)} \nonumber \\
f_{21} &= -\tfrac{1}{2} s^{3/2} ( \sqrt{16 m^2 {+} s} {+} \sqrt{s} ) E(k^2) - \tfrac{1}{2} s^{3/2} ( \sqrt{16 m^2 {+} s} {-} \sqrt{s} ) K(k^2) \\
f_{22} &= \frac{- s^{3/2} (16 m^2 {+} s) ( \sqrt{16 m^2 {+} s} {-} \sqrt{s} ) K(k^2)}{2 (1{+}2 \epsilon)} \nonumber
\end{align}

With this we may compute the differential equation, which once again is epsilon factorized and it is given by
\begin{align}
\frac{\id J_i}{\id s} = \epsilon A_{ij} J_j \; + \; \text{lower}
\end{align}
with
\begin{align}
A_{11} &= \frac{8 (12 m^2 {+} s) K(k^2) K(1{-}k^2)}{\pi \sqrt{s} (16 m^2{+}s) (\sqrt{16 m^2 {+} s}{+}\sqrt{s})} + \frac{2}{\pi s} \left( 1 {-} \frac{8 m^2}{16 m^2{+}s} {+} \frac{\sqrt{s}}{\sqrt{16 m^2 {+} s}} \right) E(k^2) E(1{-}k^2) \nonumber \\
& \;\;\; + \frac{-4 (12 m^2{+}s) K(k^2) E(1{-}k^2)}{\pi s (16 m^2 {+} s)} + \frac{-2 (\sqrt{16 m^2 {+} s} {+} \sqrt{s}) E(k^2) K(1{-}k^2)}{\pi \sqrt{s} (16 m^2 {+} s)} \nonumber \\
A_{12} &= \frac{-64 i m^2 K(1{-}k^2)^2}{\pi \sqrt{s} \sqrt{16 m^2{+}s} (\sqrt{16 m^2 {+} s} + \sqrt{s})} + \frac{i (\sqrt{16 m^2 {+} s} + \sqrt{s})^2 E(1{-}k^2)^2}{\pi s (16 m^2{+}s)} \nonumber \\
& \;\;\; + \frac{4i}{\pi s} \left( \frac{\sqrt{s}}{\sqrt{16 m^2 {+} s}} - \frac{8 m^2}{16 m^2{+}s} \right) K(1{-}k^2) E(1{-}k^2) \nonumber
\end{align}
\begin{align}
A_{21} &= \frac{i (12 m^2 {+} s) (\sqrt{16 m^2 {+} s} - \sqrt{s})^2 K(k^2)^2}{4 m^2 \pi s (16 m^2 {+} s)} + \frac{i ( \sqrt{16 m^2{+}s}-\sqrt{s} )^2 E(k^2)^2}{\pi s (16 m^2 {+} s)} \nonumber \\
& \;\;\; + \frac{-4i K(k^2) E(k^2)}{\pi s} \\
A_{22} &= \frac{2 (\sqrt{16 m^2 {+} s} - \sqrt{s})^2 K(k^2) K(1{-}k^2)}{\pi s \sqrt{16 m^2 {+} s} (\sqrt{16 m^2 {+} s} + \sqrt{s})} + \frac{-2}{s \pi} \left( 1 {-} \frac{8 m^2}{16 m^2 {+} s} {-} \frac{\sqrt{s}}{\sqrt{16 m^2 {+} s}} \right) E(k^2) E(1{-}k^2) \nonumber \\
& \;\;\; + \frac{16 m^2 K(k^2) E(1{-}k^2)}{\pi s (16 m^2{+}s)}  + \frac{-32 m^2 E(k^2) K(1{-}k^2)}{\pi s \sqrt{16 m^2 {+} s} (\sqrt{16 m^2 {+} s} + \sqrt{s})} \nonumber
\end{align}
with a similar and likewise epsilon factorized equation to be found for the other derivative $\id J/ \id m^2$.

\subsection{The two-mass elliptic sunrise}
\label{sec:twomass}

Next let us look at the two-mass elliptic sunrise (s2m). This not very well studied integral is defined as is section \ref{sec:sunrisedef} with $m_1^2 = m_2^2 = m_b^2$ and $m_3^2 = m_a^2$, making it intermediate between the same-mass elliptic sunrise of section~\ref{sec:samemass} and the three-mass elliptic sunrise of section~\ref{sec:threemass} both in terms of the number of scales and the number of master integrals. The integral family contains five master integrals; two double-tadpoles and three in the highest elliptic sector which is what will concern us here.

\begin{figure}
\centering
\includegraphics[scale=1.00]{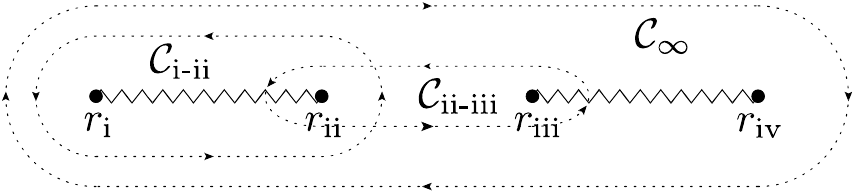}
\caption{The three independent contours for the two-mass elliptic sunrise}
\label{fig:cons2m}
\end{figure}

On the maximal cut in $d=2$ we get that the integrals in the highest sector may be written as
\begin{align}
J_i = \int_{\mathcal{C}} \frac{\hat{\phi} \, \id z}{Y} \qquad \text{with} \qquad Y = \sqrt{z \, (z {-} 4 m_b^2) \, \big( z^2 {-} 2 (m^2_a {+} s) z {+} (m^2_a {-} s)^2 \big) } 
\end{align}
We will pick the three intermediate basis integrals $I^{\text{s2m}}_{11100}$, $I^{\text{s2m}}_{21100}$, and $I^{\text{s2m}}_{111 \sminus 10}$, which correspond to the integrands
\begin{align}
\hat{\phi}_1 = 1\,, \qquad \hat{\phi}_2 = \frac{1{+}2 \epsilon}{z {-} 4 m_b^2}\,, \qquad \hat{\phi}_3 = z\,.
\end{align}
Once again we will define the master integrals we are looking for, as generic linear combinations of the intermediate basis integrals
\begin{align}
J_i = f_{i1} I^{\text{s2m}}_{11100} + f_{i2} I^{\text{s2m}}_{21100} + f_{i3} I^{\text{s2m}}_{111 \sminus 10}
\end{align}
$Y^2$ has the four roots
\begin{align}
r_{\text{i}} &= 0 \,, & r_{\text{ii}} &= (\sqrt{m_a^2} - \sqrt{s})^2 \,, & r_{\text{iii}}&= (\sqrt{m_a^2} + \sqrt{s})^2 \,, & r_{\text{iv}} &= 4 m_b^2 \,,
\end{align}
and with this we may define our three master contours
\begin{align}
\gamma_1 = \mathcal{C}_{\text{ii-iii}} \,,\qquad \gamma_2 = \mathcal{C}_{\text{i-ii}} \,,\qquad \gamma_3 = \mathcal{C}_{\infty} \,.
\end{align}
as depicted on fig.~\ref{fig:cons2m}. We will also define the abbreviations
\begin{align}
\delta_{\pm} &= \sqrt{m_a^2} \pm \sqrt{s} \,, & \mu_{\pm} &= 4 m_b^2 {-} ( \sqrt{m_a^2} {\pm} \sqrt{s} )^2 \,, \nonumber \\
\lambda_{\pm} &= m_b^2 {\pm} \sqrt{m_a^2} \sqrt{s} \,, & \rho &= 2 m_b^2 {-} m_a^2 {-} s \,,
\end{align}
and
\begin{align}
k^2 &= \frac{16 m_b^2 \sqrt{m_a^2} \sqrt{s} }{\delta_+^2 \, \mu_-}\,, & 1{-}k^2 &= \frac{\delta_-^2 \, \mu_+}{\delta_+^2 \, \mu_-}\,, & n^2 &= \frac{4 \lambda_-^2}{\rho^2}\,, & n'^2 &= \frac{\lambda_-^2}{\lambda_+^2}\,,
\end{align}
where we note the relation $(1{-}k^2)/n'^2 = 1 - k^2/n^2$. With this we can compute the needed integrals, as $g_{ij} = \int_{\gamma_j} \frac{\hat{\phi}_i \id z}{Y}$:
\begin{align}
g_{11} &= \frac{4 K(k^2)}{\delta_+ \, \sqrt{\mu_-}} &
g_{21} &= \frac{- (1{+}2 \epsilon) }{m_b^2 \sqrt{\mu_-}} \left( \frac{K(k^2)}{\delta_+} + \frac{\delta_+ E(k^2)}{\mu_+} \right) \nonumber \\
g_{12} &= \frac{-4 i K(1{-}k^2)}{\delta_+ \, \sqrt{\mu_-}} \qquad\qquad & g_{22} &= \frac{-i (1 {+} 2 \epsilon)}{\mu_+ \sqrt{\mu_-}} \left( \frac{- 4 K(1{-}k^2)}{\delta_+} + \frac{\delta_+ E(1{-}k^2)}{m_b^2} \right) \nonumber \\
g_{31} &= \frac{4 \delta_-^2}{\delta_+ \, \lambda_- \, \sqrt{\mu_-}} \left( m_b^2 K(k^2) - \frac{\lambda_+ \mu_+ \Pi(n^2,k^2)}{2 \rho} \right) \!\!\!\!\!\!\!\!\!\!\!\!\!\!\!\!\!\!\!\!\!\!\!\!\!\!\!\!\!\!\!\!\!\!\!\!\!\!\!\!\!\!\!\!\!\!\!\!\!\!\!\!\!\!\!\!\!\!\!\!\!\!\!\!\!\!\!\!\!\!\!\!\!\!\!\!\!\!\!\!\!\!\!\!\!\!\!\!\!\!\!\!\!\!\!\!\!\!\!\!\!\!\!\!\!\!\!\!\!\!\!\! && \nonumber \\
g_{32} &= i \pi + \frac{-i 4 m_b^2}{\delta_+ \lambda_- \sqrt{\mu_-}} \left( \delta_-^2 K(1{-}k^2) + \frac{2 \sqrt{m_a^2} \, \rho \, \sqrt{s} \, \Pi(n'^2, 1{-}k^2)}{\lambda_+} \right)  \!\!\!\!\!\!\!\!\!\!\!\!\!\!\!\!\!\!\!\!\!\!\!\!\!\!\!\!\!\!\!\!\!\!\!\!\!\!\!\!\!\!\!\!\!\!\!\!\!\!\!\!\!\!\!\!\!\!\!\!\!\!\!\!\!\!\!\!\!\!\!\!\!\!\!\!\!\!\!\!\!\!\!\!\!\!\!\!\!\!\!\!\!\!\!\!\!\!\!\!\!\!\!\!\!\!\!\!\!\!\!\! && \\
g_{13} &= 0 & g_{23} &= 0 \qquad\qquad\qquad\qquad\qquad\quad g_{33} = - 2 \pi i  \nonumber
\end{align}
making this the first appearance of the complete elliptic integral of the third kind $\Pi(n^2,k^2)$.

With this we may write down the period matrix $P_{ij} = f_{ik} g_{kj}$ and impose it to be diagonal $P = 2 \pi i I$ by solving for the $f_{ik}$. We get
\begin{align}
f_{11} &= \frac{-4i m_b^2 \sqrt{\mu_-} K(1{-}k^2)}{\delta_+} + i \delta_+ \sqrt{\mu_-} E(1{-}k^2) \nonumber \\
f_{21} &= \frac{-\sqrt{\mu_-} \mu_+ K(k^2)}{\delta_+} - \delta_+ \sqrt{\mu_-} E(k^2) \nonumber \\
f_{31} &= \frac{\delta_-^2 \lambda_+ \mu_p \Pi(n^2,k^2)}{\lambda_- \rho \pi} \left( \frac{4 \mu_a^2 K(1{-}k^2)}{\delta_-^2} - E(1{-}k^2) \right) - \frac{\sqrt{\mu_-} \mu_+ K(k^2)}{2 \delta_+} + \frac{\delta_-^2 m_b^2}{\lambda_-} \nonumber \\[-1.5mm]
& \;\;\; + \frac{4 m_b^2 \sqrt{m_a^2} \sqrt{s} \rho \Pi(n'^2, 1{-}k^2)}{\lambda_- \lambda_+ \pi} \left( \frac{\mu_+ K(k^2)}{\delta_+^2} K(k^2) + E(k^2) \right) - \tfrac{1}{2} \delta_+ \sqrt{\mu_-} E(k^2) \nonumber \\
f_{12} &= \frac{-4i m_b^2 \sqrt{\mu_-} \mu_p K(1{-}k^2)}{\delta_+ (1 {+} 2 \epsilon)} \qquad\qquad\qquad\qquad f_{22} = \frac{-4 m_b^2 \sqrt{\mu_-} \mu_p K(k^2)}{\delta_+ (1 {+} 2 \epsilon)} \label{eq:s2mfs} \\
f_{32} &= \frac{2 m_b^2 \mu_+}{\delta_+ (1{+}2 \epsilon)} \bigg( \frac{2 \delta_-^2 \lambda_+ \mu_+ \Pi(n^2,k^2) K(1{-}k^2)}{\delta_+ \lambda_- \rho \pi} + \frac{8 m_b^2 \rho \sqrt{m_a^2 s}\, \Pi(n'^2,1{-}k^2) K(k^2)}{\delta_+ \lambda_- \lambda_+ \pi} - \sqrt{\mu_-} K(k^2) \bigg) \nonumber \\[1mm]
f_{13} &= 0 \qquad\qquad\qquad\qquad\qquad f_{23} = 0 \qquad\qquad\qquad\qquad\qquad f_{33} = -1 \nonumber
\end{align}

Once again the differential equation matrices are epsilon factorized $\id J_i/\id x = \epsilon A^{(x)}_{ij} J_j$. The entries are in general too large to be written here, one example entry is
\begin{align}
A^{(m_b^2)}_{11} &= \frac{-12 \delta_-^2 \lambda_+ \mu_+ \Pi(n^2,k^2) K(1{-}k^2)}{\delta_+^2 \lambda_- \mu_- \rho \pi} + \frac{24 \delta_-² m_b^2 K(k^2) K(1{-}k^2)}{\delta_+^2 \lambda_- \mu_- \pi} - \frac{2 \delta_+^2 E(k^2) E(1{-}k^2)}{m_b^2 \mu_+ \pi} \nonumber \\
& \quad - \frac{8(2 \mu_+ {+} \mu_-) E(k^2)K(1{-}k^2)}{\mu_- \mu_+ \pi} - \frac{1}{m_b^2}
\end{align}
and for the full expressions see the added file.

\subsection{The three-mass elliptic sunrise}
\label{sec:threemass}

Next we will look at the generic or three-mass elliptic sunrise (s3m), exactly as it is defined in section \ref{sec:sunrisedef}. The integral family contains seven master integrals; three double-tadpoles and four in the highest, elliptic sector.

Integrals in that highest sector may, on the maximal cut and in two dimensions, be written as
\begin{align}
J_i = \int_{\mathcal{C}} \frac{\hat{\phi} \, \id z}{Y}
\end{align}
with
\begin{align}
Y = \sqrt{ \big( z^2 {-} 2 (m^2_1 {+} m^2_2) z {+} (m^2_1 {-} m^2_2)^2 \big) \big( z^2 {-} 2 (m^2_3 {+} s) z {+} (m^2_3 {-} s)^2 \big) }
\end{align}
We will pick our master integrals as linear combinations of now four intermediate basis integrals, as
\begin{align}
J_i = f_{i1} I^{\text{s3m}}_{11100} + f_{i2} I^{\text{s3m}}_{21100} + f_{i3} I^{\text{s3m}}_{111 \sminus 10} + f_{i4} I^{\text{s3m}}_{1110 \sminus 1}
\end{align}
These four intermediate basis integrals correspond to the integrands
\begin{align}
& \hat{\phi}_1 = 1\,, \qquad \hat{\phi}_2 = \frac{(1{+}2 \epsilon) (z {+} m_2^2 {-} m_1^2)}{z^2 {-} 2 (m^2_1 {+} m^2_2) z {+} (m^2_1 {-} m^2_2)^2} \,,  \qquad \hat{\phi}_3 = z\,, \nonumber \\
& \qquad\; \hat{\phi}_4 = \frac{(m_1^2{-}m_2^2) (m_3^2{-}s)}{2z} + \tfrac{1}{2} (m_1^2{+}m_2^2{+}m_3^2{+}s) - \tfrac{1}{2} z
\end{align}
where the latter, which may appear incompatible with the loop-by-loop parametrization used, has been obtained using the Passarino-Veltman inspired approach described in section 11.1 of ref.~\cite{Frellesvig:2019kgj}.

\begin{figure}
\centering
\includegraphics[scale=1.00]{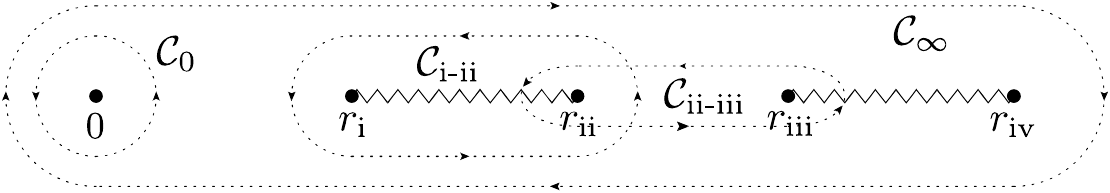}
\caption{The four independent contours for the three-mass elliptic sunrise}
\label{fig:cons3m}
\end{figure}

These four integrands are linear combination of the set
\begin{align}
& \hat{\xi}_1 = 1\,, \qquad \hat{\xi}_2 = \frac{(z {+} m_2^2 {-} m_1^2)}{z^2 {-} 2 (m^2_1 {+} m^2_2) z {+} (m^2_1 {-} m^2_2)^2} \,,  \qquad \hat{\xi}_3 = z\,, \qquad \hat{\xi}_4 = \frac{1}{z}\,,
\end{align}
of which we will compute the integrals.

The master integration contours will be chosen as
\begin{align}
\gamma_1 = \mathcal{C}_{\text{ii-iii}} \,,\qquad\quad \gamma_2 = \mathcal{C}_{\text{i-ii}} \,,\qquad\quad \gamma_3 = \mathcal{C}_{\infty} \,,\qquad\quad \gamma_4 = \mathcal{C}_{0}\,,
\end{align}
as depicted on fig.~\ref{fig:cons3m}, and the four roots of $Y^2$ are
\begin{align}
r_{\text{i}} &= \big( m_2 {-} m_1 )^2\,, & r_{\text{ii}} &= ( m_3 {-} \sqrt{s})^2\,, & r_{\text{iii}}&= ( m_3 {+} \sqrt{s})^2\,, & r_{\text{iv}} &= ( m_2 {+} m_1)^2\,.
\end{align}

We will now introduce the abbreviations
\begin{align}
\delta_{01} &= -m_1{+}m_2{+}m_3{-}\sqrt{s} \;\; & \delta_{02} &= m_1{-}m_2{+}m_3{-}\sqrt{s} \;\; & \delta_{03} &= m_1{+}m_2{-}m_3{-}\sqrt{s} \; \nonumber \\
\delta_1 &= -m_1{+}m_2{+}m_3{+}\sqrt{s} & \delta_2 &= m_1{-}m_2{+}m_3{+}\sqrt{s} & \delta_3 &= m_1{+}m_2{-}m_3{+}\sqrt{s} \nonumber \\
\delta_+ &= m_1{+}m_2{+}m_3{+}\sqrt{s} & \delta_0 &= m_1{+}m_2{+}m_3{-}\sqrt{s} & \beta &= m_1 m_2 m_3 \sqrt{s} \nonumber \\
\lambda_{1 \pm} &= m_2 m_3 \pm m_1 \sqrt{s} & \lambda_{2 \pm} &= m_1 m_3 \pm m_2 \sqrt{s} & \lambda_{3 \pm} &= m_1 m_2 \pm m_3 \sqrt{s} \\
\mu_{\pm} &= m_1 m_2 (m_3^2{-}s)^2 {\pm} (m_2^2{-}m_1^2)^2 m_3 \sqrt{s} \;\;\;\;\;\; \rho = (m_1^2{+}m_2^2) (m_3^2{-}s)^2 {-} (m_2^2{-}m_1^2)^2 (m_3^2{+}s)
\!\!\!\!\!\!\!\!\!\!\!\!\!\!\!\!\!\!\!\!\!\!\!\!\!\!\!\!\!\!\!\!\!\!\!\!\!\!\!\!\!\!\!\!\!\!\!\!\!\!\!\!\!\!\!\!\!\!\!\!\!\!\!\!\!\!\!\!\!\!\!\!\!\!\!\!\!\!\!\!\!\!\!\!\!\!\!\!\!\!\!\!\!\!\!\!\!\!\!\!\!\!\!\!\!\!\!\!\!\!\!\!\!\!\!\!\!\!\!\!\!\!\!\!\!\!\!\!\!\!\!\!\!\!\!\!\!\!\!\!\!\!\!\!\!\!\!\!\!\! && \nonumber \\
q &= (m_3^2{-}m_2^2)^2 {+} 2(m_3^2{+}m_2^2)(m_1^2{-}s) {+} s^2{+}2m_1^2s{-}3m_1^4 \!\!\!\!\!\!\!\!\!\!\!\!\!\!\!\!\!\!\!\!\!\!\!\!\!\!\!\!\!\!\!\!\!\!\!\!\!\!\!\!\!\!\!\!\!\!\!\!\!\!\!\!\!\!\!\!\!\!\!\!\!\!\!\!\!\!\!\!\!\!\!\!\!\!\!\!\!\!\!\!\!\!\!\!\!\!\!\!\
&&& \psi_{\pm} &= (m_1^2 {\pm} m_2^2 {-} m_3^2 {-} s) \nonumber
\end{align}
and
\begin{align}
k^2 &= \frac{16 \beta}{\delta_0 \delta_1 \delta_2 \delta_3} & n^2 &= \frac{4 \lambda_{3-}^2}{\psi_+^2} & \tilde{n}^2 &= \frac{4 \mu_-^2}{\rho^2} \nonumber \\
(1{-}k^2) &= \frac{\delta_+ \delta_{01} \delta_{02} \delta_{03}}{\delta_0 \delta_1 \delta_2 \delta_3} & n'^2 &= \frac{\lambda_{3-}^2}{\lambda_{3+}^2} & \tilde{n}'^2 &= \frac{\mu_-^2}{\mu_+^2}
\end{align}
where again we have the relation $(1{-}k^2)/n'^2 = 1 - k^2/n^2$ and likewise for $\tilde{n}$. 
With this we may write the integrals $h_{ij} = \int_{\gamma_j} \xi_i/Y$:
\begin{align}
h_{11} &= \frac{4 K(k^2)}{\sqrt{\delta_0 \delta_1 \delta_2 \delta_3}} \qquad\qquad h_{21} = \frac{-1}{\sqrt{\delta_0 \delta_1 \delta_2 \delta_3} \, m_1^2} \left( K(k^2) + \frac{q E(k^2)}{\delta_+ \delta_{01} \delta_{02} \delta_{03}} \right) \nonumber \\
h_{31} &= \frac{2}{\sqrt{\delta_0 \delta_1 \delta_2 \delta_3} \, \lambda_{3-}} \left( 2 \lambda_{1-} \lambda_{2-} K(k^2) - \frac{\delta_+ \delta_{01} \delta_{02} \delta_{03} \lambda_{3+} \Pi(n^2,k^2)}{\psi_+} \right) \nonumber \\
h_{41} &= \frac{2}{\sqrt{\delta_0 \delta_1 \delta_2 \delta_3} \, \mu_{-}} \left( 2 \lambda_{1-} \lambda_{2-} K(k^2) - \frac{\delta_+ \delta_{01} \delta_{02} \delta_{03} \mu_+ \Pi(\tilde{n}^2,k^2)}{\rho} \right) \nonumber \\
h_{12} &= \frac{-4i K(1{-}k^2)}{\sqrt{\delta_0 \delta_1 \delta_2 \delta_3}} \qquad\qquad h_{22} = \frac{-i \big( q E(1{-}k^2) + 4 ( 2 \beta {+} m_1^2 \psi_-) K(1{-}k^2) \big)}{\sqrt{\delta_0 \delta_1 \delta_2 \delta_3} \, \delta_+ \delta_{01} \delta_{02} \delta_{03} m_1^2} \nonumber \\
h_{32} &= \frac{-4i}{\sqrt{\delta_0 \delta_1 \delta_2 \delta_3} \, \lambda_{3-}} \left( \lambda_{1-} \lambda_{2-} K(1{-}k^2) + \frac{2 \beta \psi_+ \Pi(n'^2,1{-}k^2)}{\lambda_{3+}} \right) + i \pi \\
h_{42} &= \frac{-4i}{\sqrt{\delta_0 \delta_1 \delta_2 \delta_3} \, \mu_{-}} \left( \lambda_{1-} \lambda_{2-} K(1{-}k^2) + \frac{2 \beta \rho \Pi(\tilde{n}'^2,1{-}k^2)}{\mu_+} \right) - \frac{i \pi}{(m_2^2{-}m_1^2) (m_3^2{-}s)} \nonumber \\
h_{13} &= h_{23} = h_{43} = h_{14} = h_{24} = h_{34} = 0 \qquad h_{33} = -2 \pi i \qquad h_{44} = \frac{2 \pi i}{(m_2^2{-}m_1^2) (m_3^2{-}s)} \nonumber
\end{align}
We may then form the integrals of the $\hat{\phi}$ integrands $g_{ij} = \int_{\gamma_j} \phi_i/Y$ as combinations of the above, and then form the period matrix $P_{ij} = f_{ik} g_{kj}$. Fixing it to be diagonal $P = 2 \pi i I$ will fix all 16 $f_{ij}$. We will not list them all here, two examples are
\begin{align}
f_{11} &= i \sqrt{\delta_0 \delta_1 \delta_2 \delta_3} \left( \frac{4 (2 \beta {+} m_1^2 \psi_-)}{q} K(1{-}k^2) + E(1{-}k^2) \right) \,, \nonumber \\
f_{31} &= \left( \frac{\delta_+ \delta_{01} \delta_{02} \delta_{03} }{q} K(k^2) + E(k^2) \right) \left( \frac{4 \beta \psi_+}{\lambda_{3-} \lambda_{3+} \pi} \Pi(n'^2,1{-}k^2) - \frac{\sqrt{\delta_0 \delta_1 \delta_2 \delta_3}}{2} \right) \\
& \;\;\; + \frac{- \delta_+ \delta_{01} \delta_{02} \delta_{03} \lambda_{3+}}{\lambda_{3-} \psi_+ \pi} \left( \frac{4 (2 \beta {+} m_1^2 \psi_-)}{q} K(1{-}k^2) + E(1{-}k^2) \right) \Pi(n^2,k^2) + \frac{\lambda_{1-} \lambda_{2-}}{\lambda_{3-}}. \nonumber
\end{align}

With this choice we get the differential equations in epsilon-factorized form $\id J_i/\id x = \epsilon A^{(x)}_{ij} J_j$ for all the kinematic variables $x$. The matrix entries are too large to be written here.

\subsection*{The added expressions}

In many of the examples done in this section, the result were deemed too large to be written out in full. For that reason a file {\tt expressions.m} has been added to the arXiv version of this paper, which contains all the expressions in Mathematica format. The expressions are named for instance {\tt Ass2m} for the matrix $A^{(s)}_{\text{s2m}}$ or {\tt frulesnpt} for replacement rules for the $f_{ij}$ in the case of the non-planar double triangle. To obtain the full list of expressions defined in the file one might use the Mathematica command {\tt Names["Global`*"]} after reading it in.

\section{Discussion}
\label{sec:discussion}

\subsection{Form and integration of the differential equations}

One topic that has been avoided so far in this paper, is how to solve the differential equations once the epsilon-factorized form has been obtained. For Feynman integrals in the traditional canonical form it is often straight forward to integrate them up order by order in the epsilon expansion, giving results in the function class of generalized polylogarithms~\cite{Goncharov:1998kja}, (at least in the case where all square-roots can be rationalized, which is however known to not always be possible~\cite{Brown:2020rda}). Many attempts have been made to extend this function class to elliptic cases and beyond~\cite{brown2011multiple, Adams:2014vja, Broedel:2017kkb, Adams:2017ejb}, yet none of those seem directly applicable to the form of the differential equations derived in this paper. Further investigations will be needed into how to best proceed analytically from the differential equations found here.

If, however, one is satisfied with a numerical approach, the development of techniques to numerically integrate this type of differential equations is developing quickly~\cite{Boughezal:2007ny, Czakon:2007qi, Aglietti:2007as, Lee:2017qql, Mandal:2018cdj, Moriello:2019yhu, Hidding:2020ytt}. Not all of these approaches require the differential equations to be epsilon factorized, but they all benefit from from it and work faster in that case, so even in the worst case scenario the techniques developed here can be used to speed up the numerics.

We do however notice one property that the differential equations found on the previous pages share with those in traditional canonical form and which might cause optimism for the prospect of analytical integration: They are free of higher poles at singular points including the point at infinity. This property is not obvious, it might appear as if for instance the $A_{11}$-element of the differential equation matrix found for the same-mass sunrise of eqs.~\eqref{eq:difeqs1m} would have a higher pole in $\sqrt{s} = -m$. But that is not the case. Changing variable to $y = \sqrt{s}+m$, setting $m = 1$, and expanding, that matrix element becomes
\begin{align}
A^{\text{s1m};(y)}_{11} &= \frac{1}{\pi} \bigg( \big( 6 {-} 2 \pi i {+} 6 \log(y/4) \big) \frac{1}{y} + \tfrac{1}{4} \big( 15 {-} 7 \pi i {+} ( 33 {-} 6 i \pi ) \log(y/4) {+} 18 \log^2(y/4) \big) \nonumber \\
& + \tfrac{1}{64} \big( 273 {-} 39 i \pi {+} ( 309 {-} 60 i \pi ) \log(y/4) {+} 180 \log^2(y/4) \big) y + \mathcal{O}(y^2) \bigg)
\end{align}
and we see that no higher pole in $y$ is present, the candidate having canceled with a term from the expansion of the elliptic integrals.

\subsection{Freedom in basis choice}
\label{sec:freedom}

There is some amount of freedom in the algorithm discussed on the previous pages. This freedom corresponds to performing row operation on the period matrix. For instance one might define the master contours $\gamma_i$ as some linear combination of the ones used in this paper. In all the elliptic examples we chose $\gamma_1 = \mathcal{C}_{\text{ii-iii}}$ and $\gamma_2 = \mathcal{C}_{\text{i-ii}}$, but one might equally well have chosen for instance $\gamma_2 = \mathcal{C}_{\text{iii-iv}}$ which would yield a slightly different set of integrals in the end, and also there is nothing preventing the choice of some $\mathbb{C}$-linear combination of contours, such as $\gamma = \tfrac{5}{7} \mathcal{C}_{\text{ii-iii}} + \tfrac{8i}{13} \mathcal{C}_{\text{i-ii}}$. Yet these kinds of redefinitions seem to provide no simplification of the integrals or their differential equations. Likewise one might take linear combinations of the resulting integrals, since if the set $J$ has epsilon factorized differential equations so will a $\mathbb{C}$-linear combination $\tilde{J} = B J$ with $B_{ij} \in \mathbb{C}$. For example for the case of the two-mass elliptic sunrise, it seems from eqs.~\eqref{eq:s2mfs} as if it would provide a slight simplification to perform the replacement $J_3^{\text{s2m}} \rightarrow \tilde{J}_3^{\text{s2m}} = J_3^{\text{s2m}} - \tfrac{1}{2} J_2^{\text{s2m}}$. 

Yet as mentioned in the introduction, there are previous examples in the literature in which elliptic Feynman integrals have been put into a form that allows for epsilon factorized differential equations, and which are not equal, or in a form related through the above mentioned transformations, to the expressions found here.
In particular the same-mass sunrise of section \ref{sec:samemass} has been discussed in ref.~\cite{Adams:2018yfj}, and the three-mass sunrise of section \ref{sec:threemass} has been discussed in ref.~\cite{Bogner:2019lfa}. It must be admitted that the integrals found in those references are ``nicer'' than those found with the algorithm outlined here. In particular the matrix of $f_{ij}$ coefficients are in both cases found to be something more akin to a lower triangular form and the differential equation matrix is nicer as well as we will see. Focusing on the same-mass sunrise, the form found in ref.~\cite{Adams:2018yfj} has $f_{12}=0$ which means that the master integrals there may be written
\begin{align}
\tilde{J}^{\text{s1m}}_1 &= c f_{11} I^{\text{s1m}}_{11100} \nonumber \\
\tilde{J}^{\text{s1m}}_2 &= c (f_{21} I^{\text{s1m}}_{11100} + f_{22} I^{\text{s1m}}_{21100}) \label{eq:s1mwein}
\end{align}
where the values are found to
\begin{align}
f_{11} &= \frac{- \epsilon \pi R}{4 K(k^2)} \nonumber \\
f_{21} &= \frac{R E(k^2)}{4 \pi} + \frac{\big( 3 (m^2{-}s) (9 m^2{-}s) + 2 \epsilon (45 m^4 {-} 30 m^2 s {+} s^2) \big) K(k^2)}{12 \pi R} \label{eq:fsolwein}
\end{align} \begin{align}
f_{22} &= \frac{m^2 R (\sqrt{m^2}{-}\sqrt{s}) (3 \sqrt{m^2}{+}\sqrt{s}) K(k^2)}{\pi (\sqrt{m^2}{+}\sqrt{s})^2} \nonumber
\end{align}
with the definitions of $k$ and $R$ being as given in sec. \ref{sec:samemass}. We see that these expressions are quite different from what was found in eqs.~\eqref{eq:fsols1m}. Aside from having three coefficients rather than four, eqs.~\eqref{eq:fsolwein} has simpler kinematics dependence, and only complete elliptic integrals of $k^2$ appear where eqs.~\eqref{eq:fsols1m} also contained complete elliptic integrals of argument $(1{-}k^2)$. On the other hand the $\pi$ and $\epsilon$ dependence is simpler in eqs.~\eqref{eq:fsols1m}.

From this we may compute the differential equations. They may be written as $\id \tilde{J} / \id s = \epsilon A \tilde{J}^{\text{s1m}}$ with
\begin{align}
A_{11} &= \frac{9 m^4 {+} 10 m^2 s {-} 3 s^2}{2 s (9 m^2{-}s)(m^2{-}s)} \qquad\qquad\qquad\qquad\qquad\quad\;\; A_{12} = \frac{3 \pi^2 (\sqrt{m^2}{+}\sqrt{s})^2}{4 s (\sqrt{m^2}{-}\sqrt{s}) (3 \sqrt{m^2}{+}\sqrt{s}) K(k^2)^2} \nonumber \\
A_{21} &= \frac{(3 m^2 {+} s)^4 K(k^2)^2}{3 \pi^2 s (\sqrt{m^2}{-}\sqrt{s}) (3 \sqrt{m^2}{-}\sqrt{s})^2 (\sqrt{m^2}{+}\sqrt{s})^4 (3\sqrt{m^2}{+}\sqrt{s})} \qquad\; A_{22} = \frac{9 m^4{+}10 m^2 s {-} 3 s^3}{2 s (9 m^2{-}s) (m^2{-}s)}
\label{eq:difeqwein}
\end{align}
which is simpler than the expressions in eqs.~\eqref{eq:difeqs1m} by a substantial amount, we see for instance that the only elliptic integral appearing is $K(k^2)$, and that the diagonal entries are free of even this. On the other hand the $K(k^2)$ appears both in numerator and denominator, where in eqs.~\eqref{eq:difeqs1m} the elliptic integrals appeared in the numerator only.

We may compute the period matrix of this example. The result is $P_{ij}$ with
\begin{align}
P_{11} &= -\pi \eps \qquad\qquad\qquad\qquad\qquad\qquad P_{12} = \frac{i \pi \epsilon K(1{-}k^2)}{K(k^2)} \nonumber \\
P_{21} &= \frac{2 \eps K(k^2)}{\pi} \left( \frac{2 (3 m^2{-}s) (3 m^2 {+} s) K(k^2)}{3 R^2} - E(k^2) \right) \\
P_{22} &= \frac{-i}{2} - \frac{2 i \epsilon K(k^2)}{\pi} \left( \frac{(3 \sqrt{m^2} {+} \sqrt{s}) ( 3 m^3 {-} 9 m^2 \sqrt{s} {-} 3 m s {+} s^{3/2} )}{R^2} K(1{-}k^2) + E(1{-}k^2) \right) \nonumber
\end{align}
and in addition it is worth noticing that the determinant of this matrix is given as
\begin{align}
\text{det}(P) &= \tfrac{1}{2} i \pi \epsilon (1 {+} 2 \epsilon)
\end{align}

We see that $P_{11}$ and $\text{det}(P)$ both are constants, in the sense that they are free of any kinematic dependence including through elliptic integrals. This may be used as a guideline for obtaining an $\epsilon$-factorized form alternative to the one discussed in this paper. We do an example of this in Appendix~\ref{app:nptagain}, but it is not clear if this can be generalized to elliptic sectors with more than two master integrals.

It would be great if some principle could be found to a priori generate the transformation between the forms discussed in this paper, and forms similar to the one discussed above.

\subsection{The number of cycles}
\label{sec:numberofcycles}

In principle there is a one-to-one correspondence between the number of master integrals and the number of independent cycles~\cite{Lee:2013hzt, Bitoun:2017nre, Frellesvig:2019uqt}. This correspondence is utilized by the Lee-Pomeransky criterion~\cite{Lee:2013hzt}, which states that the number of master integrals $\nu$ is given as
\begin{align}
\nu &= \text{number of solutions to ``} \omega = 0 \text{''} \qquad \text{where} \qquad \omega := \id \text{log}(u)
\end{align}
We can test this for the examples discussed in secs.~\ref{sec:motivation} and \ref{sec:examples} of this paper. The result is that the numbers agree in all the cases, except for the same-mass sunrise of sec.~\ref{sec:samemass} and the nonplanar double triangle of sec.~\ref{sec:npt}, which are the two cases in which an elliptic sector has two master integrals. For the case of the same-mass sunrise, this ``miscounting'' is well known, and is discussed in further detail in refs.~\cite{Frellesvig:2019kgj, Weinzierl:2020xyy} where it is shown to be caused by peculiarities in the interplay between the loop-by-loop Baikov parametrization and the maximal cut.

Focusing on the same-mass sunrise it might seem as if a valid third master integral would be $I_{111 \sminus 10}^{\text{s1m}}$ corresponding to $\hat{\phi}_3{=}z$, with a corresponding third master contour surrounding the pole at infinity $\gamma_3 {=} \mathcal{C}_{\infty}$ making the set of cycles look as in the case of two-mass elliptic sunrise of fig.~\ref{fig:cons2m}. Yet we know from IBP relations that
\begin{align}
I_{111 \sminus 10}^{\text{s1m}} &= \frac{3 m^2 {+} s}{3} I_{11100}^{\text{s1m}} \, + \, \text{lower} \label{eq:s1mibp}
\end{align}
and it might be instructive to see how this relation is realized over the basic cycles. Using the $\hat{\phi}_i$ and $\gamma_j$ from sec.~\ref{sec:samemass}, and again defining $g_{ij} = \int_{\gamma_j} \hat{\phi}_i \id z /Y$, we get
\begin{align}
g_{31} &= \int_{\gamma_1} \! \frac{\hat{\phi}_3 \id z}{Y} = 2 (\sqrt{m^2}{-}\sqrt{s}) \left( 2 \sqrt{m^2} K(k^2) - (3\sqrt{m^2}{+}\sqrt{s}) \Pi(n^2,k^2) \right) \! \big/ R \nonumber \\
g_{32} &= \int_{\gamma_2} \! \frac{\hat{\phi}_3 \id z}{Y} = -4i \left( \sqrt{m^2} (\sqrt{m^2}{-}\sqrt{s}) K(1{-}k^2) + 2 \sqrt{m^2} \sqrt{s} \Pi(n'^2,1{-}k^2) \right) \! \big/ R + i \pi \\
g_{13} &= \int_{\gamma_3} \! \frac{\hat{\phi}_1 \id z}{Y} = 0 \qquad\qquad g_{23} = \int_{\gamma_3} \! \frac{\hat{\phi}_2 \id z}{Y} = 0 \qquad\qquad g_{33} = \int_{\gamma_3} \! \frac{\hat{\phi}_3 \id z}{Y} = -2 \pi i \nonumber
\end{align}
where $k^2$ and $R$ are as defined in sec.~\ref{sec:samemass}, and where
\begin{align}
n^2 = \frac{4 m^2}{(\sqrt{m^2}{+}\sqrt{s})^2} \qquad\quad \text{and} \qquad\quad n'^2 = \frac{(\sqrt{m^2}{-}\sqrt{s})^2}{(\sqrt{m^2}{+}\sqrt{s})^2}
\end{align}
If eq.~\eqref{eq:s1mibp} were to hold on the contour $\gamma_j$ it would mean that $g_{3j} = ((3 m^2 {+} s)/3) g_{1j}$, so let us investigate the status of that relation on each contour.

Starting by $\gamma_1$, it might seem unlikely for the relation to hold since $g_{31}$ contains $\Pi(n^2,k^2)$, which $g_{11}$ does not. But it does hold exactly, due to the relation
\begin{align}
\Pi \! \left( \frac{4}{(1{+}x)^2} , \frac{-16 x}{(x{-}3) (1{+}x)^3} \right) &= \frac{2x}{3(x{-}1)} K \! \left( \frac{-16 x}{(x{-}3) (1{+}x)^3} \right)
\end{align}
(valid for $x \in [-1,1]$). On the other hand for contour $\gamma_3$ the relation clearly does not hold since $g_{13}=0$, while $g_{33}=-2\pi i$ which is then the amount with which the relation is broken. At the last contour $\gamma_2$ we get using a similar relation for a special point of $\Pi(n^2,k^2)$, that
\begin{align}
g_{32} - \frac{3 m^2 {+} s}{3} g_{12} &= \frac{2 \pi i}{3}
\end{align}
so we see that when the peculiarities of the loop-by-loop Baikov parametrization ruins the counting of master contours, relations such as eq.~\eqref{eq:s1mibp} break only by a factor proportional to $2 \pi i$.

For the nonplanar double triangle of sec.~\ref{sec:npt} the behaviour is extremely similar. There the extra IBP-derived integral relation relating an integral with a pole at infinity to the original two, is
\begin{align}
I_{111111 \sminus 1}^{\text{npt}} &= \frac{-s}{2} I_{1111110}^{\text{npt}} \, + \, \text{lower}
\end{align}
and that relation holds exactly on contour $\gamma_1$ and is broken with factors proportional to $2 \pi i$ on $\gamma_2$ and a potential $\gamma_3$ surrounding a pole at infinity. The main difference is that the $\Pi(n^2,k^2)$ special value relation that makes this explicit is even simpler in that case.

\subsection{Freedom in intermediate basis choice}
\label{sec:freedomintermediate}

One might consider if the intermediate bases chosen in the examples discussed in secs.~\ref{sec:motivation} and \ref{sec:examples} have some special property that allows for the algorithm to go through that would not be shared with any arbitrary basis choice. The answer to this is mostly no, any set of valid master integrals might be chosen as the intermediate basis. Let us illustrate that with an example, that of the two-mass elliptic sunrise of sec.~\ref{sec:twomass}. Instead of the choice of ($I^{\text{s2m}}_{11100}$, $I^{\text{s2m}}_{21100}$, $I^{\text{s2m}}_{111 \sminus 10}$) made in sec.~\ref{sec:twomass}, one might instead pick for instance the set ($I^{\text{s2m}}_{11100}$, $I^{\text{s2m}}_{111 \sminus 10}$, $I^{\text{s2m}}_{111 \sminus 20}$) which is the choice naturally made by Kira.

With this one can go through the same procedure as before, define
\begin{align}
\tilde{J}_i = \tilde{f}_{i1} I^{\text{s2m}}_{11100} + \tilde{f}_{i2} I^{\text{s2m}}_{111 \sminus 10} + \tilde{f}_{i3} I^{\text{s2m}}_{111 \sminus 20}
\end{align}
compute the period matrix, and impose it to be proportional to $I$ thereby fixing the $\tilde{f}_{ij}$. The resulting $\tilde{J}_i$ equal those from sec.~\ref{sec:twomass} up to integrals in lower sectors as well as terms of order $\epsilon$:
\begin{align}
\tilde{J}_i = J_i + \mathcal{O}(\epsilon) + \text{lower}
\end{align}
Additionally the new integrals have epsilon factorized differential equations $\id \tilde{J}_i/\id x = \epsilon \tilde{A}^{(x)}_{ij} \tilde{J}_j$ in each of the three kinematic variables $x \in \{m_a^2, m_b^2, s\}$. The expressions for $\tilde{f}$ and the $\tilde{A}^{(x)}$ can be found in the file added to the arXiv version of this paper.

There are however cases in which not any choice of intermediate basis is equally convenient. This is true in the cases that have the ``miscounting'' of independent cycles discussed in sec.~\ref{sec:numberofcycles}, such as the same-mass elliptic sunrise. There we saw that a counting of independent cycles seems to include a third master contour surrounding a pole at infinity $\mathcal{C}_{\infty}$, and that the relation reducing integrals with support there to integrals with support only on the original two $\mathcal{C}_{\text{ii-iii}}$ and $\mathcal{C}_{\text{i-ii}}$ is realized only up to factors proportional to $2 \pi i$. For the same-mass elliptic sunrise, ($I^{\text{s1m}}_{11100}$, $I^{\text{s1m}}_{111 \sminus 20}$) which is chosen as the default by Kira would be a valid alternative to the intermediate basis discussed in section~\ref{sec:samemass}. But since the integrand of $I^{\text{s1m}}_{111 \sminus 20}$ has support on $\mathcal{C}_{\infty}$, such a factor proportional to $2 \pi i$ has to be removed by hand before a basis with epsilon factorized differential equations is obtained.

\subsection{Further discussion, open questions, and conclusions}

In all the examples in this paper, we only discussed the highest sector for the integral families we looked at, which allowed us to only approach the system on its maximal cut. Yet for practical applications one would need the whole system of differential equations to be in epsilon-factorized form, not just the highest sector. Presumably the approach discussed here could be generalized away from the maximal cut, by introducing a multivariate notion of master contours, and indeed this is the approach taken in the prescriptive unitarity scheme~\cite{Bourjaily:2017wjl, Bourjaily:2021vyj}. Yet to bring subsectors into epsilon-factorized form, I do not believe this is needed. It will be enough to fit coefficients of lower sector integrals using an ansatz with its free coefficients fixed from imposing the epsilon-factorized property of the differential equations, starting from the highest subsectors going down, along the lines of the method proposed in ref.~\cite{Gehrmann:2014bfa} for the non-elliptic case.

Likewise the discussion here was limited to the case where all subsectors were free of elliptic contributions and could be put into canonical form in the traditional sense. For integrals for which that is not the case (such as the kite integrals that will be obtained by promoting $D_4$ and $D_5$ to genuine propagators for the elliptic sunrise integrals) presumably a similar approach can be taken, but that is a matter for further study.

The canonical form proposed in ref.~\cite{Henn:2013pwa} not only implied an epsilon-factorized form for the differential equations of eq.~\eqref{eq:epsfac}, but further imply that the distinct differential equations for each kinematic variable are joined in one differential of dlog-form $\id J = \epsilon (\id M) J$ where $M$ is a matrix containing only logarithms of algebraic functions of the kinematic variables. An obvious question is if the differential equations found here can likewise be unified into one differential form $\id J = \epsilon (\id M) J$, but where $M$ obviously no longer will contain solely logarithms. In the polylogarithmic case, the arguments of those logarithms are the symbol letters~\cite{Goncharov:2010jf, Duhr:2012fh} of the problem, so an obvious open question is if the entries of $M$ in the elliptic case, if it even exists, will be related to the entries of the elliptic generalization of the coaction~\cite{Broedel:2018iwv} in a similar straight forward fashion. For the polylogarithmic case, the fact that the algorithm discussed in this paper works in the first place, may be seen as following from the existence of an iterated dlog representation for the whole integral~\cite{Arkani-Hamed:2010pyv, WasserMSc, Herrmann:2019upk, Dlapa:2021qsl}, and the question of how such a representation might generalize to the elliptic case is presumably also linked to the above.

The answers to such questions will help clarify what the ``correct'' generalization of terms such as \textit{pure}~\cite{Broedel:2018qkq}, \textit{canonical}, \textit{weight}, etc. that have proven extremely helpful in the understanding of polylogarithmic Feynman integrals, will be to the elliptic case and beyond, if such generalizations can even be made in a useful way. I did not attempt to make any such generalizations in this work, but I hope that my approach contributes with insights that may help bring us to a point where such a step might be taken.

The algorithm presented in this paper is able to systematically bring differential equations for elliptic Feynman integrals into a form for which the differential equations are epsilon factorized. We demonstrated this for a number of examples of varying complexity, and it is the hope that this development will be one step on the way towards bringing the understanding of elliptic Feynman integrals to the same level as the polylogarithmic case.

\subsubsection*{Acknowledgments}

The author would like to thank Pierpaolo Mastrolia and the rest of the \textit{intersection} group in Padova, as well as Cristian Vergu and the rest of the NBIA \textit{amplitudes} group, for many helpful discussions. Particular thanks go to Pierpaolo Mastrolia, Manoj Mandal, Matt von Hippel, and Stefan Weinzierl, for carefully reading through the manuscript at the draft stage and providing helpful feedback.

This project has received funding from the European Union's Horizon 2020 research and innovation program under the Marie Sk{\l}odowska-Curie grant agreement No. 847523 'INTERACTIONS'. The work has been partially supported by a Carlsberg Foundation Reintegration Fellowship.

\appendix

\section{A different form for the elliptic nonplanar double triangle}
\label{app:nptagain}

In the example discussed in section \ref{sec:freedom}, we saw that a valid form for the equal mass sunrise integral~\cite{Adams:2018yfj} resulting in an $\epsilon$-factorized differential equation could be made without the diagonal period matrix otherwise discussed in this paper. That example had $f_{12}=0$ and $P_{11}$ and $\text{det}(P)$ both being constants. Using these properties as a guideline, we were able to find an epsilon factorized basis for the nonplanar double triangle of section~\ref{sec:npt}, which has two master integrals in the elliptic sector similarly to the equal mass sunrise.

We write the general form
\begin{align}
\tilde{J}^{\text{npt}}_1 &= c f_{11} I^{\text{npt}}_{1111110} \nonumber \\
\tilde{J}^{\text{npt}}_2 &= c (f_{21} I^{\text{npt}}_{1111110} + f_{22} I^{\text{npt}}_{2111110}) \label{eq:nptnew}
\end{align}
where we have used $f_{12}=0$. Imposing $P_{11}$ and $\text{det}(P)$ being constants puts two constraints on the remaining $f_{ij}$ while the third may be fitted by requiring the differential equation to be $\epsilon$-factorized.
We find the values
\begin{align}
f_{11} &= \frac{\epsilon \, s^{3/2} (\sqrt{16 m^2{+}s}{+}\sqrt{s})}{4 K(k^2)} \nonumber \\
f_{21} &= - s^{3/2} \left( (\sqrt{16 m^2 {+} s} {+} \sqrt{s}) E(k^2) + (\sqrt{16 m^2{+}s}{-}\sqrt{s}) \left( 1 + \epsilon \frac{24 m^2{+}s}{m^2} \right) K(k^2) \right) \\
f_{22} &= - s^{3/2} (16 m^2 {+} s) (\sqrt{16 m^2{+}s}{-}\sqrt{s}) K(k^2) \nonumber
\end{align}
which corresponds to the period matrix (given the two contours defined in eqs.~\eqref{eq:gammanpt})
\begin{align}
P_{11} &= 2 \epsilon \qquad\qquad\qquad\qquad\qquad P_{12} = - 2 i \epsilon \frac{K(1{-}k^2)}{K(k^2)} \nonumber \\
P_{21} &= 2 \epsilon K(k^2) \left( 8 E(k^2) - \frac{(16 m^2{+}s) (\sqrt{16 m^2{+}s} {-} \sqrt{s})}{m^2 (\sqrt{16 m^2{+}s} {+} \sqrt{s})} K(k^2) \right) \\
P_{22} &= 4 i \pi + 2 i \epsilon K(k^2) \left( 8 E(1{-}k^2) + \frac{\big( (8m^2{+}s) \sqrt{16 m^2{+}s} - (24 m^2{+}s) \sqrt{s} \big)}{m^2 (\sqrt{s}{+}\sqrt{16 m^2{+}s})} K(1{-}k^2) \right) \nonumber
\end{align}
with the property
\begin{align}
\text{det}(P) &= 8 i \pi \eps (1 {+} 2 \eps)
\end{align}
and we see the two requirements imposed above, being apparent.

The corresponding $s$-differential equation $\id \tilde{J}^{\text{npt}}_i / \id s = \epsilon \tilde{A}_{ij} \tilde{J}^{\text{npt}}_j$ has
\begin{align}
\tilde{A}_{11} &= \frac{- (8 m^2 {+} s)}{s (16 m^2 {+} s)} \qquad\qquad\qquad\qquad\;\;
\tilde{A}_{12} = \frac{(\sqrt{16 m^2{+}s}{+}\sqrt{s})^2}{8 \sqrt{2} s (16 m^2{+}s) K(k^2)^2} \nonumber \\
\tilde{A}_{21} &= \frac{8 \sqrt{2} (8 m^2{+}s)^2 K(k^2)^2}{s (16 m^2{+}s) (\sqrt{16 m^2{+}s} {+} \sqrt{s})^2} \qquad\qquad\qquad \tilde{A}_{22} = \frac{- (8 m^2 {+} s)}{s (16 m^2 {+} s)}
\end{align}
with a similar differential equation in the other variable $m^2$. This form of the differential equation resembles that given in eq.~\eqref{eq:difeqwein}, and as such it will likely be suitable for the integration approach proposed in ref.~\cite{Adams:2018yfj}.

It would be interesting to generalize these considerations to a more general elliptic case, or preferably to find a principle that allows for reliably generating a transformation that can map between the forms discussed in the bulk of this paper and forms similar to the above, but to what extend that is possible is a question for the future.

\bibliographystyle{JHEP}
\bibliography{biblio}

\end{document}